\documentclass[11pt,eqs]{article}
\usepackage{latexsym,amsmath}
\usepackage{amssymb}
\usepackage{hyperref}
\usepackage{multicol}
\usepackage[thinlines]{easytable}

\def\cleq{\setcounter{equation}{0}}
\title{Courant algebroid without constraints on fluxes on its Dirac structures}
\author{I. Ivani\v sevi\'c \thanks{e-mail: ivanisevic@ipb.ac.rs} and B. Sazdovi\'c
\thanks{e-mail: sazdovic@ipb.ac.rs}\\
{\it Institute of Physics, University of Belgrade}\\
{\it Pregrevica 118, 11080 Belgrade, Serbia}}

\begin{document}
\maketitle
\begin{abstract}
We examine the standard Courant bracket and its extensions, defined by twists with different $O(D,D)$ transformations relevant to string theory. We analyze Dirac structures on these Courant algebroids and derive the constraints they impose on fluxes. In the end, we show that the Courant algebroid simultaneously twisted by both $B$ and $\theta$ is characterized by Dirac structures with no restrictions on fluxes.
\end{abstract}

\section{Introduction}
\cleq

The treatment of the elements of smooth section of tangent bundle (vector fields) and cotangent bundle (1-forms) as components of a single element (generalized vector) is fundamental of generalized geometry development \cite{gualtieri, GCY}. It encompasses many well studied geometries, such as Poisson manifolds, symplectic structures, and Kähler manifolds. The smooth section of the generalized tangent bundle is equipped with the Courant bracket and Courant algebroid, that respectively generalize the well known Lie bracket and Lie algebroid. 

The manifestation of these mathematical structures in string theory becomes evident when one considers T-duality, a transformation connecting equivalent string theories defined in different geometries or even topologies \cite{tdual, tdual1, tdual2, buscher, buscher1}. Primarily, the Courant bracket governs algebra of string diffeomorphisms and local gauge transformations, which are known to be mutually related by T-duality \cite{doucou, dualsim}. Secondarily, string fluxes appear as structure functions in the algebra of various twisted Courant brackets \cite{royt, royt1, cdual, CBTh}. The fluxes are essential for string stabilization of moduli \cite{flux1, flux2, flux3}, and they are known to be related to string non-commutativity and non-associativity \cite{NC-string, NA-string, fluxNO}. Moreover, to each Courant algebroid corresponds a Courant $\sigma$-model suitable for description of topological branes \cite{aksz}. This action is of the AKSZ type, and its complete BV-BRST formulation can be described in terms of $L_{\infty}$ algebras \cite{L-infty}.

From the viewpoint of string theory exploring Dirac structures has proven to be intriguing. These are isotropic subspaces of maximal dimension where section elements are closed under the Courant (or other Courant algebroid) bracket \cite{courant,  dirak, dirak1}. Dirac structures define the integrability conditions of the almost complex structures. In theory of open bosonic strings, a D-branes can be identified as a leaf of foliated structure, which are known to be integrable submanifolds \cite{courant, dirak2}. In fact, D-branes can be described exactly as Dirac structures, such that a scalar field and a gauge field on the D-brane are treated on an equal footing \cite{asakawa}.

The essential feature of Dirac structures is that on them Leibniz and Jacobi identities are satisfied for the Courant bracket. This motivated authors in \cite{nilm, bulk} to search for the Dirac structures on nilmanifolds that contain all fluxes, in search for non-geometric backgrounds, containing all fluxes and not being T-dual to any geometric one. It has been demonstrated that one can construct the extension of the Courant bracket that contains both $H$ and $R$ flux, such that all fluxes can indeed exist on its Dirac structures, though certain restrictions on fluxes needed to be imposed. 

The natural question to ask is: ''Is it possible to construct a Courant algebroid such that all fluxes can exist on its Dirac structures with no restrictions imposed on them?'' \footnote{All fluxes must satisfy Bianchi identities, such as the ones obtained in \cite{flux3}, which are the identities intrinsic to the way how fluxes are defined. When we pose this question, we are specifically asking whether there exist Dirac structures with fluxes that impose no additional restrictions on the fields used to define these fluxes.} In order to answer this question affirmatively, we extend our previous research with regards to different geometrical structures from the symmetry generator algebra. When generators of general coordinate transformations and local gauge transformations are combined into a single generator, it has the form of the inner product on generalized tangent bundle with a generalized vector as its symmetry parameter. The Courant bracket arises in the Poisson bracket algebra between such extended generators \cite{cdual}. 

This approach is convenient as it is possible to act with an arbitrary $O(D, D)$ transformation on the basis in which the generator is expressed, obtaining the same generator in the basis of non-canonical currents. The algebra of such a generator produces twisted Courant brackets, where twisting is governed exactly by the $O(D, D)$ element used for the basis transformation. A plethora of twisted Courant brackets can be generated this way, including the well known $B$-twisted Courant bracket \cite{courant1} (featuring $H$-flux) and the $\theta$-twisted Courant bracket (featuring $Q$ and $R$ fluxes), that are themselves known to be related by T-duality \cite{crdual}. 

Successive twisting by $B$ and then by $\theta$ leads to the twisted Courant bracket that features all fluxes, also known as the Roytenberg bracket \cite{royt}. However, the $B$- and $\theta$-transformations do not commute and the underlining generator in which algebra this bracket is obtained is not invariant under T-duality. To achieve this, one can perform the twisting by $B$ and $\theta$ at the same time, obtaining the Courant bracket twisted simultaneously by $B$ and $\theta$ \cite{CBTh}. All fluxes characterized by this bracket are themselves complicated hyperbolic function of both $B$ and $\theta$ fields \cite{fluksevi-rad}. As it will be pointed out in this paper, the Courant bracket twisted by both $B$ and $\theta$ defines the Courant algebroid such that all fluxes can exist on its Dirac structures without any constraints imposed on them.

In the first chapter, we deal with the mathematical preliminaries. We procede by considering the so called standard Courant algebroid, defined with the Courant bracket, $O(D, D)$ invariant inner product, and the projection to the tangent bundle as its anchor. We show that symplectic and Poisson manifolds are its Dirac structures. Though these results are already well known \cite{gualtieri, courant}, we present them for completeness and in order to introduce the unfamiliar reader with the simplest case. 

In the third chapter, we show how every element of the $O(D,D)$ group naturally defines an isomorphism between the standard Courant algebroid and the one associated with the twisted Courant bracket, where the twist is determined by the same $O(D,D)$ element. The appropriate transformation of the anchor ensures the manifest satisfaction of all Courant algebroid compatibility conditions. Though Courant algebroids related with $O(D,D)$ are isomorphic, the particular realisations of algebroids and their respective Dirac structures are different. This can affect different physical properties, such as fluxes, interpretations of dualities and symmetries, and (non)-geometric nature of the background. 

In the proceeding two chapters, we illustrate the characteristics of some well known Courant algebroids and their associated Dirac structures, including the case of $B$-twisted Courant bracket and $\theta$-twisted Courant bracket. In the former case, the restrictions on $H$-flux on its Dirac structures are lifted, while in the latter case, the restrictions on the $R$-flux on its Dirac structures are lifted. Afterwards, we consider the Courant bracket twisted firstly by $B$, and then by $\theta$. Unlike the previous twisted brackets, in this one all fluxes are present. However, we show that restrictions on fluxes on Dirac structures still apply.
 
Lastly, we consider the Courant bracket twisted at the same time by $B$ and $\theta$ and its respective Courant algebroid. On its associated Dirac structures, we demonstrate that all fluxes can exist. This way, we obtain the most important result that has motivated our paper. 

\section{Action, mathematical preliminaries and notations}
\cleq

Consider a closed bosonic string moving in the background characterized with the symmetric metric tensor $G_{\mu \nu}$ and the antisymmetric Kalb-Ramond field $B_{\mu \nu}$. Its motion is described by action 
\cite{action, regal}
\begin{equation} \label{eq:action}
S = \int d\sigma d\tau {\cal L} = \int d\sigma d\tau  \partial_+ x^\mu \Pi_{+\mu \nu} \partial_\nu x^\nu \, , \quad \Pi_{\pm \mu \nu} = B_{\mu \nu} \pm \frac{1}{2}G_{\mu \nu} \, ,
\end{equation}
where $x^\mu$ are the coordinates on the worldsheet parametrized with one spacelike parameter $\sigma$ and one timelike parameter $\tau$, and
\begin{equation}
\partial_{\pm} x^\mu  = \dot{x}^\mu \pm x^{\prime \mu} \, , \quad \dot{x}^\mu = \partial_\tau x^\mu , \ x^{\prime \mu} = \partial_\sigma x^\mu \, .
\end{equation}
It is straightforward to obtain canonical momenta by varying the Lagrangian with respect to the time-like coordinate derivative
\begin{equation} \label{eq:pi-can}
\pi_\mu = \frac{\partial {\cal L}}{\partial \dot{x}^\mu}= \kappa G_{\mu \nu} \dot{x}^{\nu} - 2\kappa B_{\mu \nu} x^{\prime \nu} \, ,
\end{equation}
allowing us to obtain the Hamiltonian by
\begin{equation} \label{eq:Hcan-def}
{\cal H}_{\cal C} = \pi_\mu \dot{x}^\mu - {\cal L} = \frac{1}{2\kappa} \pi_\mu (G^{-1})^{\mu \nu} \pi_\nu -2 x^{\prime \mu} B_{\mu \nu} (G^{-1})^{\nu \rho} \pi_\rho + \frac{\kappa}{2} x^{\prime \mu} G^E_{\mu \nu} x^{\prime \nu} \, . 
\end{equation}
We assume that the canonical variables satisfy the usual Poisson bracket relations
\begin{equation} \label{eq:PBR}
\{ x^{\mu} (\sigma), \pi_\nu (\bar{\sigma}) \} = \delta^\mu_{\ \nu} \delta(\sigma - \bar{\sigma}) \, ,
\end{equation}
with other brackets being zero.

The symmetries of the space-time consists of the general coordinate transformations \cite{doucou, dualsim}
\begin{equation} \label{eq:deltaxi}
\delta_\xi x^\mu = -\xi^\mu \, ,
\end{equation}
under which the background fields transform by
\begin{eqnarray}\label{eq:dif-GB}
\delta_\xi G_{\mu \nu} = \xi^\rho \partial_\rho G_{\mu \nu} + \partial_\mu \xi^\rho G_{\rho \nu}  + \partial_\nu \xi^\rho G_{\rho \mu}  \,  , \nonumber \\
\delta_\xi B_{\mu \nu} = \xi ^\rho \partial_\rho B_{\mu \nu} + \partial_\mu \xi^\rho B_{\rho \nu}  - \partial_\nu \xi^\rho B_{\rho \mu } \, .
\end{eqnarray}
In a canonical approach the generator of transformations (\ref{eq:deltaxi}) is given by 
\begin{eqnarray}\label{eq:Gxi}
{\cal G}_\xi = \int d \sigma \, \xi^\mu \pi_\mu \, ,
\end{eqnarray}
where $\pi_\mu$ is canonical momentum (\ref{eq:pi-can}). We have
\begin{eqnarray}\label{eq:Ict2}
\delta_\xi x^\mu = \{{\cal G}_\xi , x^\mu \}   =   - \xi^\mu  \, .
\end{eqnarray}

Besides the general coordinate transformations, string backgrounds are invariant under the so called local gauge transformations \cite{doucou, dualsim}
\begin{equation} \label{eq:lg-GB}
\delta_{\lambda} G_{\mu \nu} = 0 \, , \quad \delta_{\lambda} B_{\mu \nu} = \partial_\mu \lambda_{\nu} - \partial_\nu \lambda_\mu \, .
\end{equation}
The generator of such transformation is given by
\begin{equation} \label{eq:Gl}
{\cal G}_{\lambda} = \int d\sigma \kappa \lambda_\mu x^{\prime \mu} \, .
\end{equation}

Physically equivalent descriptions are connected using the T-duality \cite{tdual, tdual1, tdual2, buscher, buscher1}. The T-dual background fields are 
\begin{equation}
{}^\star B^{\mu \nu} = \frac{\kappa}{2}\theta^{\mu \nu} , \, \quad {}^\star G^{\mu \nu} = G^E_{\mu \nu} \, , 
\end{equation}
where $\theta$ is the so called non-commutativity parameter, given by
\begin{equation} \label{eq:thetadef}
\theta^{\mu \nu} = -\frac{2}{\kappa} (G_E^{-1} B G^{-1})^{\mu \nu} = -\frac{2}{\kappa} (G^{-1} B G_E^{-1})^{\mu \nu} \, ,
\end{equation}
and $G_E$ is the effective metric, given by
\begin{equation} \label{eq:Geff}
G^E_{\mu \nu} = G_{\mu \nu}  - 4 B_{\mu \rho}(G^{-1})^{\rho \sigma} B_{\sigma \nu} \, .
\end{equation}
Both the non-commutativity parameter and the effective metric are features of the effective open sting theory obtained for the solution of boundary conditions. The local gauge transformations are found to be T-dual to the general coordinate transformations \cite{doucou, dualsim}.

\subsection{Lie derivative}

The Lie derivative determines the change of any tensor field along the parametrized flow defined by a vector field. Specifically, the Lie derivative of the tensor field T of type (p, q) with respect to the vector field $\xi$ is another tensor field ${\cal L}_{\xi} T$ of type (p, q) with components
\begin{eqnarray}\label{eq:def-lie-der} 
{\cal L}_{\xi} T^{\mu_1 \mu_2 ...  \mu_p}_{\nu_1 \nu_2 ... \nu_q} = \xi^\rho \partial_\rho T^{\mu_1 \mu_2 ...  \mu_p}_{\nu_1 \nu_2 ... \nu_q}-\sum_{i=0}^p T^{\mu_1  ... \rho \hat{\mu}_i ... \mu_p}_{\nu_1 \nu_2 ... \nu_q}\partial_\rho \xi^{\mu_i}+\sum_{i=1}^q T^{\mu_1  ...  \mu_p}_{\nu_1  ... \rho \hat{\nu}_i ...\nu_q}\partial_{\nu_i} \xi^{\rho} \, ,
\end{eqnarray}
where $\hat{\nu}_i$ denotes omission of such index, e.g. $T^{\mu_1  \mu_2 ...  \mu_p}_{\nu_1   ... \rho \hat{\nu}_q}\partial_{\nu_q} \xi^{\rho } = T^{\mu_1 \mu_2 ...  \mu_p}_{\nu_1 \nu_2 ... \rho}\partial_{\nu_q} \xi^\rho $. 
In the local coordinate basis, the action of Lie derivative on vectors and functions is given by
\begin{equation}
({\cal L}_{\xi} \eta )^\mu = \xi^\nu \partial_\nu \eta^\mu- \eta^\nu \partial_\nu \xi^\mu \, , \quad {\cal L}_{\xi} f = \xi^\mu \partial_\mu f \, ,
\end{equation}
where $\eta$ is a smooth vector field, and $f$ is a smooth function. The action of Lie bracket between two vector fields $\xi_1$ and $\xi_2$ on a smooth function is defined through the relation 
\begin{equation} \label{eq:lie-br-def}
[\xi_1, \xi_2]_L f = ({\cal L}_{\xi_1} \xi_2) f \, .
\end{equation}
It is a well known fact that for each smooth function $f$ on a manifold ${\cal M}$ the Lie bracket satisfies the Leibniz rule 
\begin{equation} \label{eq:lleibniz}
[ \xi_1, f \xi_2 ]_L = f [\xi_1, \xi_2]_L + ({\cal L }_{\xi_1} f)\ \xi_2 \, ,
\end{equation}
as well as the Jacobi identity
\begin{equation} \label{eq:JAC}
[\xi_1, [\xi_2, \xi_3]_L ]_L + [\xi_2, [\xi_3, \xi_1]_L ]_L +[\xi_3, [\xi_1, \xi_2]_L ]_L = 0 \, . 
\end{equation}
It is exactly the Lie bracket that governs the Poisson bracket algebra of a string general coordinate transformations 
\begin{equation}
\{ {\cal G}_{\xi_1}, {\cal G}_{\xi_2}\} = - {\cal G}_{[\xi_1, \xi_2]_L} \, .
\end{equation}

Apart from vectors and functions, the Lie derivative can also act on differential $p$-forms $\lambda$ by
\begin{equation} \label{eq:Lxil}
{\cal L}_\xi  \lambda = d i_\xi \lambda + i_\xi d \lambda \, ,
\end{equation}
where $i_\xi$ is the interior derivative, and $d$ is the exterior derivative. The interior derivative is a degree -1 derivation of the algebra of differential forms, given by
\begin{equation} \label{eq:int-der}
(i_\xi \lambda) (\xi_1, \xi_2, ... \xi_{p-1}) = \lambda (\xi, \xi_1, \xi_2, ... \xi_{p-1}) \, ,
\end{equation}
where $\xi_1$, ... , $\xi_p$ are vectors. In case of 1-form $\lambda$, we have
\begin{equation} \label{eq:ixidef}
i_\xi \lambda = \xi^\mu \lambda_\mu \, .
\end{equation}

The exterior derivative is a degree 1 derivation of the algebra of differential forms. On function it acts as a differential operator
\begin{equation}
d f (\xi) = {\cal L}_\xi f  \Leftrightarrow  (df)_\mu = \partial_\mu f \, .
\end{equation}
In case of an arbitrary $p$-form $\lambda$, the exterior derivative is defined by
\begin{eqnarray} \label{eq:ext-der}
d \lambda (\xi_1, ... ,\xi_p) &=& \sum_{i=1}^p (-1)^{i+1} \xi_i \Big( \lambda(\xi_1, ..., \hat{\xi}_i, ..., \xi_p) \Big) \\ \notag
&&+ \sum_{i<j} (-1)^{i+j} \lambda([\xi_i, \xi_j]_L, \xi_1, ... ,\hat{\xi}_i, ... ,\hat{\xi}_j , ... ,\xi_p) \, ,
\end{eqnarray}
where $\hat{\xi}_i$ denotes the emission of $\xi_i$, and $\xi(\lambda) = \xi^\mu \partial_\mu \lambda$. In case of a 1-form $\lambda$, the above expression reduces to
\begin{equation}
(d \lambda)_{\mu \nu} = \partial_\mu \lambda_\nu - \partial_\nu \lambda_\mu \, .
\end{equation}

\subsection{Lie algebroid}

Let us now consider a more general vector bundle $V$ over a manifold ${\cal M}$ with a bracket on its space of smooth sections, together with a vector bundle map $\rho : V \to T{\cal M}$, called the anchor.  The structure $(V, [,], \rho)$ is called the Lie algebroid \cite{liealg, liealg1} if 
\begin{equation} \label{eq:lie-alg1}
\rho [\xi_1, \xi_2] = [\rho (\xi_1), \rho (\xi_2)]_L \, ,
\end{equation}
and the bracket satisfies the Jacobi identity (\ref{eq:JAC}), and the generalization of the Leibnitz rule
\begin{equation} \label{eq:lie-alg2}
[ \xi_1, f \xi_2 ] = f [\xi_1, \xi_2] + ({\cal L}_{\rho(\xi_1)} f)\ \xi_2 \, .
\end{equation}
In a similar manner, one can define the exterior derivative for any Lie algebroid by \cite{LA-geometry}
\begin{eqnarray} \label{eq:ext-der-rho}
d \lambda (\xi_1, ... ,\xi_p) &=& \sum_{i=1}^p (-1)^{i+1} \rho(\xi_i) \Big( \lambda(\xi_1, ... ,\hat{\xi}_i, ... ,\xi_p) \Big) \\ \notag
&&+ \sum_{i<j} (-1)^{i+j} \lambda([\xi_i, \xi_j], \xi_1, ... ,\hat{\xi}_i, ... ,\hat{\xi}_j,  ..., \xi_p) \, ,
\end{eqnarray}
where $\lambda$ are elements of the smooth sections of the $n$-th exterior power of the dual bundle $V^{\star}$, i.e. $n$-covectors. They are the Lie algebroid analogs of the differential $n$-forms.

\subsubsection{Examples of Lie algebroids}

To illustrate the notion of Lie algebroid, let us give a couple of well known examples. The first and the simplest one consists of the tangent bundle $T{\cal M}$ with the usual Lie bracket (\ref{eq:lie-br-def}) and the identity operator $\text{Id}$ as its anchor $(T{\cal M}, [,]_L, \text{Id})$. The dual bundle is cotangent bundle $T^{\star}{\cal M}$, and on the smooth section of its exterior product one can define the usual exterior derivative (\ref{eq:ext-der}) as the Lie algebroid derivative (\ref{eq:ext-der-rho}).

To see some less trivial example, let us consider the cotangent bundle $T^{\star} {\cal M}$ with a bi-vector $\theta$ defining a morphism $\hat{\theta}$ by
\begin{equation} \label{eq:theta-morph}
\hat{\theta} (\lambda_1)\lambda_2 = \theta(\lambda_1, \lambda_2) = \theta^{\mu \nu} \lambda_{1\mu} \lambda_{2\nu} \Rightarrow (\hat{\theta} (\lambda_1))^\mu =  \lambda_{1\nu}  \theta^{\nu \mu}  \, .
\end{equation}
Moreover, let us assume that the bi-vector $\theta$ is Poisson, i.e. its Schouten-Nijenhuis bracket \cite{SNB} is zero $[\theta, \theta]_{S} = 0 \, $, where 
\begin{equation} \label{eq:SNb}
\left. [\theta, \theta]_S \right| ^{\mu \nu \rho} = \epsilon^{\mu \nu \rho}_{\alpha \beta \gamma} \theta^{ \alpha \sigma} \partial_\sigma \theta^{\beta \gamma} \, ,
\end{equation}
and the alternator is given by
\begin{equation}
\epsilon^{\mu \nu \rho}_{\alpha \beta \gamma} = 
\begin{vmatrix}
\delta^\mu_\alpha & \delta^\nu_\beta & \delta^\rho_\gamma \\ 
\delta^\nu_\alpha & \delta^\rho_\beta & \delta^\mu_\gamma \\
\delta^\rho_\alpha & \delta^\mu_\beta & \delta^\nu_\gamma
\end{vmatrix}\, .
\end{equation}
In order for the morphism $\hat{\theta}$ to define an anchor of Lie algebroid, we need to have a bracket that satisfies
\begin{equation}
\hat{\theta} \Big( [\lambda_1, \lambda_2] \Big) = [\hat{\theta}(\lambda_1), \hat{\theta}(\lambda_2)]_L \, ,
\end{equation}
which is satisfied for the Koszul bracket, given by \cite{koszul}
\begin{equation} \label{eq:koszul-def}
[\lambda_1, \lambda_2]_{\theta} = {\cal L}_{\hat{\theta} (\lambda_1)}\lambda_2-{\cal L}_{\hat{\theta} (\lambda_2)}\lambda_1-d (\theta(\lambda_1, \lambda_2)) \, ,
\end{equation}
which in some local coordinates is given by
\begin{equation} \label{eq:koszul-def-coord}
\Big([\lambda_1, \lambda_2]_{\theta} \Big)_\mu = \theta^{\rho \nu} (\lambda_{1\rho} \partial_\nu \lambda_{2\mu}-\lambda_{2\rho} \partial_\nu \lambda_{1\mu}) + \partial_\mu \theta^{\nu \rho} \lambda_{1\nu} \lambda_{2\rho} \, .
\end{equation}
The structure $\Big(T^\star {\cal M}, [,]_{\theta}, \hat{\theta}\Big)$ is the Lie algebroid, for any Poisson bi-vector $\theta$. In case of the bi-vector $\theta$ not being a Poisson one, the Koszul bracket can still be defined, though it is no longer a Lie algebroid bracket.

Since this Lie algebroid is defined on cotangent bundle, it defines the derivation on the tangent bundle. From (\ref{eq:ext-der-rho}), we can explicitly write the action of exterior derivative on functions and vectors
\begin{equation} \label{eq:ext-der-theta}
(d_{\theta} f)^\mu=\theta^{\mu \nu}\partial_\nu f \, , \quad (d_{\theta} \xi)^{\mu \nu}  = \theta^{\mu \rho}\partial_\rho \xi^\nu - \theta^{\nu \rho}\partial_\rho \xi^\mu - \xi^\rho \partial_\rho \theta^{\mu \nu} \, ,
\end{equation}
while its action on multi-vectors can be as easily obtained. The generalized formula for the exterior derivative can also be written in terms of Schouten-Nijenhuis bracket \cite{flux3}
\begin{equation}
d_{\theta} = [\theta, . ]_{S} \, .
\end{equation}

\subsection{Generalized tangent bundle}

The generalized tangent bundle \cite{gualtieri} is the direct sum of a tangent and cotangent bundle over a manifold. Its elements are generalized vectors decomposable into
\begin{equation} \label{eq:Lxi}
\Lambda^M = \begin{pmatrix}
\xi^\mu \\
\lambda_\mu \\
\end{pmatrix} \, ,
\end{equation}
where $\xi^\mu$ represents the vector components from the smooth section of the tangent bundle $T{\cal M}$, and $\lambda_\mu$ represents the 1-form components from the smooth section of the cotangent bundle $T^{\star}{\cal M}$. The generalized tangent bundle is naturally equipped with the inner product 
\begin{equation} \label{eq:skalproizvod}
\langle \Lambda_1,\Lambda_2\rangle  = (\Lambda_1^T)^M \eta_{MN} \Lambda_2^N   \, ,
\end{equation}
where $\eta_{MN}$ is the $O(D,D)$ invariant metric, given by
\begin{equation} \label{eq:Omegadef}
\eta_{MN} = 
\begin{pmatrix}
0 & 1 \\
1 & 0 
\end{pmatrix} \, .
\end{equation}
The inner product can also be expressed in terms of the components as
\begin{equation} \label{eq:innerpr}
\langle \Lambda_1,\Lambda_2\rangle = \langle \xi_1 \oplus \lambda_1 ,\xi_2 \oplus \lambda_2 \rangle = i_{\xi_1} \lambda_2 + i_{\xi_2} \lambda_1 \, ,
\end{equation}
where $i_\xi $ represents the interior derivative (\ref{eq:ixidef}).

These mathematical structures are convenient to describe string symmetries. The generator of both general coordinate and local gauge transformation is given by 
\begin{equation}\label{eq:gltilde}
{\cal G}_{\xi \oplus \lambda}=\int_0^{2\pi} d\sigma\Big[\xi^\mu\pi_\mu+ \lambda_\mu \kappa x^{\prime\mu} \Big] \, ,
\end{equation}
which can be simply rewritten as the inner product in the space of generalized vectors
\begin{equation} \label{eq:Ggen}
{\cal G}_{\Lambda} = \int d\sigma \langle \Lambda,X \rangle \, ,
\end{equation}
where $X$ is a double canonical variable, given by
\begin{equation} \label{eq:Xdouble}
X^M = \begin{pmatrix}
\kappa x^{\prime \mu} \\
\pi_\mu \\
\end{pmatrix}\, .
\end{equation}

Assuming the standard canonical coordinate-momentum relations, it was shown that parameters of the symmetry generators in the Poisson bracket algebra of their respective generators $(\ref{eq:gltilde})$ give rise to the Courant bracket \cite{doucou, cdual}
\begin{equation} \label{eq:GGcourant}
\Big\{ {\cal G}_{\Lambda_1} , \, {\cal G}_{\Lambda_2}\Big\} = - {\cal G}_{[\Lambda_1,\Lambda_2]_{\cal C}} \, ,
\end{equation}
which is the skew-symmetric bracket which generalizes the Lie bracket on a smooth section of the generalized tangent bundle, defined by \cite{courant}
\begin{equation}
[\Lambda_1,\Lambda_2]_{\cal C} = \Lambda \Leftrightarrow [\xi_1 \oplus \lambda_1, \xi_2 \oplus \lambda_2]_{\cal C} = \xi  \oplus \lambda \, ,
\end{equation}
where
\begin{equation} \label{eq:xiC}
\xi = [ \xi_1, \xi_2]_L \, ,
\end{equation}
and
\begin{equation} \label{eq:lambdaC}
\lambda = {\cal L}_{\xi_1} \lambda_2 - {\cal L}_{\xi_2} \lambda_1 - \frac{1}{2} d (i_{\xi_1} \lambda_2 - i_{\xi_2} \lambda_1) \, .
\end{equation}
In local coordinate basis, expression (\ref{eq:xiC}) and (\ref{eq:lambdaC}) can respectively be written as
\begin{equation} \label{eq:xicou}
\xi^\mu = \xi_1^\nu \partial_\nu \xi_2^\mu - \xi_2^\nu \partial_\nu \xi_1^\mu
\end{equation}
and
\begin{equation} \label{eq:Lcou}
\lambda_\mu = \xi_1^\nu (\partial_\nu \lambda_{2 \mu} - \partial_\mu \lambda_{2 \nu}) - \xi_2^\nu (\partial_\nu \lambda_{1 \mu} - \partial_\mu \lambda_{1 \nu})+\frac{1}{2} \partial_\mu (\xi_1 \lambda_2- \xi_2 \lambda_1 ) \, .
\end{equation}
It is clear that on the tangent bundle, the Courant bracket reduces to the Lie bracket.

\subsection{Courant algebroid}

The Courant algebroid is the structure on the generalized tangent bundle analogous to the Lie algebroid. It was firstly introduced in \cite{courant1} as the Drinfel’d double \cite{drinfeld1, drinfeld2} of a Lie bialgebroid - a structure that contains two Lie algebroid structures on two dual vector bundles satisfying certain compatiblity conditions between its brackets (see \cite{liealg1} for details, and \cite{drinfeld3} for some recent developments in its generalizations).

To define the Courant algebroid, one introduces a vector bundle $E$, together with the non-degenerate inner product $\langle , \rangle$ and a skew-symmetric bracket $[ , ]$ on a smooth section of a vector bundle $E$, with a smooth bundle map $\rho : E \to T M$  as an anchor \footnote{There is an alternative (though equivalent) way to define the Courant algebroid in which instead of antisymmetric bracket we have a bilinear operator that is not necessarely antisymmetric, but satisfies the Jacobi identity \cite{royt1}.}. Then, it is possible to define the differential operator on $C^{\infty}(E)$ by 
\begin{equation} \label{eq:calDdef}
\langle {\cal D} f , \Lambda \rangle =  {\cal L}_{\rho(\Lambda)} f \, .
\end{equation}
The structure $\Big( E, \langle, \rangle, [, ], \rho \Big)$ is called the Courant algebroid if it satisfies the following compatibility relations between its elements
\begin{eqnarray} 
&& \rho [\Lambda_1, \Lambda_2] = [\rho (\Lambda_1), \rho(\Lambda_2) ]_L \, , \label{eq:CAdef1} \\ 
&& [\Lambda_1, f \Lambda_2] = f [\Lambda_1, \Lambda_2] + ({\cal L}_{\rho(\Lambda_1)} f) \Lambda_2   - \frac{1}{2}\langle \Lambda_1, \Lambda_2 \rangle {\cal D}f  \, , \label{eq:CAdef2} \\ 
&&  {\cal L}_{\rho(\Lambda_1)}\langle \Lambda_2, \Lambda_3 \rangle = \langle [\Lambda_1, \Lambda_2] +\frac{1}{2} {\cal D} \langle \Lambda_1, \Lambda_2 \rangle , \Lambda_3 \rangle + \langle \Lambda_2,  [\Lambda_1, \Lambda_3] + \frac{1}{2}{\cal D} \langle \Lambda_1, \Lambda_3 \rangle \rangle \, , \label{eq:CAdef3} \\ 
&&  \langle {\cal D} f, {\cal D} g\rangle = 0  \, ,  \label{eq:CAdef4} \\ 
&& \text{Jac}(\Lambda_1, \Lambda_2, \Lambda_3) = {\cal D} \text{Nij}(\Lambda_1, \Lambda_2, \Lambda_3)  \, , \label{eq:CAdef5}
\end{eqnarray}
for all $\Lambda_1$, $\Lambda_2$, $\Lambda_3$ from smooth section of a vector bundle $V$, and for all smooth functions $f$ and $g$ on the manifold. The Jacobiator $\text{Jac}$ is given by
\begin{equation} \label{eq:Jacobiator}
\text{Jac}(\Lambda_1,\Lambda_2,\Lambda_3) = [[\Lambda_1,\Lambda_2 ]\, ,\Lambda_3] + [[\Lambda_2,\Lambda_3 ]\, ,\Lambda_1] + [[\Lambda_3,\Lambda_1 ]\, ,\Lambda_2]   \, ,
\end{equation} 
and $\text{Nij}$ is the Nijenhuis operator, given by
\begin{equation} \label{eq:Nij}
\text{Nij}(\Lambda_1,\Lambda_2,\Lambda_3) =\frac{1}{6}\Big( \langle [\Lambda_1,\Lambda_2],\Lambda_3 \rangle +\langle [\Lambda_2,\Lambda_3],\Lambda_1 \rangle +\langle [\Lambda_3,\Lambda_1],\Lambda_2 \rangle \Big) \, .
\end{equation}

The most well known example of the Courant algebroid, what we will denote as the standard Courant algebroid, consists of the generalized tangent bundle $T{\cal M} \oplus T^{\star} {\cal M}$ with the Courant bracket, and the natural projection $\pi$ as its anchor, given by 
\begin{equation}\label{eq:piL}
\pi (\Lambda) = \pi (\xi \oplus \lambda) = \xi  \, ,
\end{equation}
which in the matrix notation can be expressed as
\begin{equation} \label{eq:pimatrix}
\pi = \begin{pmatrix}
1 & 0 \\
0 & 0 
\end{pmatrix} \, .
\end{equation}
The differential operator is just the exterior derivative $d$. It is well known in literature (see for example \cite{gualtieri} for proof) that the compatibility conditions (\ref{eq:CAdef1}) - (\ref{eq:CAdef5}) are satisfied. Throughout the paper other examples of Courant algebroids will be presented.

\subsection{Dirac structures}

Both of the Leibniz and Jacobi identity are violated with the term in the form of the inner product of two generalized vectors. It is clear that the sub-bundles so that the inner product of vectors in its section is zero will satisfy both the Jacobi identity and the Leibniz rule.
The sub-bundles on which 
\begin{equation}
\langle \Lambda_1, \Lambda_2 \rangle = 0  \, ,
\end{equation}
for each generalized vectors $\Lambda_1$ and $\Lambda_2$ are called isotropic sub-bundles. The isotropic spaces with the maximal dimension that are closed under the skew-symmetric bracket are called Dirac structures \cite{courant}. 

For any generalized vector, there are two natural projections on isotropic spaces. For a 2-form $B$, one can define
\begin{eqnarray}\label{eq:isoB}
{\cal I}_B (\Lambda) = \xi^\mu \oplus 2 B_{\mu \nu} \xi^\nu \, ,   \qquad B_{\mu \nu} = - B_{\nu \mu} \, ,
\end{eqnarray}
while for a bi-vector $\theta$, one can define
\begin{eqnarray}\label{eq:isoth}
{\cal I}_{\theta} (\Lambda) = \kappa  \theta^{\mu \nu} \lambda_{\nu} \oplus \lambda_{\mu} \, ,   \qquad \theta^{\mu \nu} = - \theta^{\nu \mu} \, .
\end{eqnarray}
The sub-bundle in the form (\ref{eq:isoB}) we will refer to as the graph of $B$ over the tangent bundle, while the sub-bundle in the form (\ref{eq:isoth}) we will refer to as the graph of $\theta$ over the cotangent bundle. In \cite{gualtieri} (Proposition 2.6, page 7), it has been demonstrated that all maximally isotropic suspaces are in the form of graphs (\ref{eq:isoB}) and (\ref{eq:isoth}). In case when isotropic subspaces are in addition closed under the appropriate algebroid bracket, they are Dirac structures. 

Dirac structures can define various different geometries, which can be illustrated on the example of standard Courant algebroid. Let us firstly consider the condition for the isotropic space in the image of projection ${\cal I}_B$ to be a Dirac structures. Substituting the relations between the vector and 1-form components (\ref{eq:isoB}) in (\ref{eq:xiC}) and (\ref{eq:lambdaC}), one obtains
\begin{equation}
[ \xi_1^{\mu} \oplus 2B_{\mu \rho} \xi_1^{\rho}, \ \xi_2^{\nu} \oplus 2B_{\nu \sigma} \xi_2^{\sigma} ]_{\cal C} = \xi^\mu \oplus \lambda_{\mu} \, ,
\end{equation}
where
\begin{equation} \label{eq:xicou}
\xi^\mu = \xi_1^\nu \partial_\nu \xi_2^\mu - \xi_2^\nu \partial_\nu \xi_1^\mu \, , \notag
\end{equation}
and
\begin{equation}
\lambda_{\mu} = 2 B_{\mu \rho} (\xi^{\nu}_1 \partial_{\nu} \xi^{\rho}_2 - \xi^{\nu}_2 \partial_{\nu} \xi^{\rho}_1) - 2 B_{\mu \nu \rho} \xi^{\nu}_1 \xi_2^{\rho} \, , 
\end{equation}
where $B_{\mu \nu \rho}$ is the Kalb-Ramond field strength (or $H$-flux), given by
\begin{equation} \label{eq:bmnr}
B_{\mu \nu \rho} = \partial_\mu B_{\nu \rho} + \partial_\nu B_{\rho \mu} + \partial_\rho B_{\mu \nu} \, .
\end{equation}
Due to this term, ${\cal I}_B (\Lambda)$ is a Dirac structure only in a special case of a closed form 2-form 
\begin{equation} \label{eq:IBIB}
[{\cal I}_B(\Lambda_1), {\cal I}_B(\Lambda_2)]_{\cal C} = {\cal I}_B \Big( [\Lambda_1,\Lambda_2]_{\cal C} \Big) \, ,\ d B = 0 \, .
\end{equation}
We can identify this as a symplectic structure for case of a non-degenerate $B$, or presymplectic structure in a more general case.

In the case of ${\cal I}_{\theta}$ (\ref{eq:isoth}), we have
\begin{equation}
[\theta^{\mu \rho} \lambda_{1\rho} \oplus \lambda_{1\mu},\ \theta^{\nu \sigma} \lambda_{2\sigma} \oplus \lambda_{2\nu}]_{\cal C} =\xi^\mu \oplus \lambda_{\mu} \, ,
\end{equation}
where 
\begin{equation}
\xi^\mu = \theta^{\mu \sigma} \theta^{\nu \rho} (\lambda_{1\rho}\partial_\nu \lambda_{2\sigma}-\lambda_{2\rho}\partial_\nu \lambda_{1\sigma}) + (\theta^{\nu \rho}\partial_\nu \theta^{\mu \sigma}-\theta^{\nu \sigma}\partial_\nu \theta^{\mu \rho})\lambda_{1\rho}\lambda_{2\sigma} = \theta^{\mu \sigma} \lambda_{\sigma} +R^{\mu \rho \sigma} \lambda_{1\rho} \lambda_{2\sigma} \, ,
\end{equation}
and
\begin{equation} \label{eq:lam-st-calg}
\lambda_{\mu} = \theta^{\nu \rho} (\lambda_{1\rho} \partial_{\nu} \lambda_{2\mu}-\lambda_{2\rho} \partial_{\nu} \lambda_{1\mu}) + \partial_\mu \theta^{\nu \rho} \lambda_{1\rho} \lambda_{2\nu} \, ,
\end{equation}
where $R$ is the non-geometric flux, given by
\begin{equation} \label{eq:nongeomflux}
 R^{\mu \nu \rho} = \theta^{\mu \sigma} \partial_\sigma \theta^{\nu \rho} + \theta^{\nu \sigma} \partial_\sigma \theta^{\rho \mu} +\theta^{\rho\sigma} \partial_\sigma \theta^{\mu \nu} \, .
\end{equation}
The $R$ flux can be written as Schouten-Nijenhuis bracket  $2R =  [\theta, \theta]_S$ (\ref{eq:SNb}), so we see that the subspace ${\cal I}_{\theta}$ with $\theta$ being the Poisson bi-vector represents a Dirac structure
\begin{equation}
[{\cal I}_\theta(\Lambda_1), {\cal I}_\theta(\Lambda_2)]_{\cal C} = {\cal I}_\theta\Big( [\Lambda_1,\Lambda_2]_{\cal C} \Big)\, ,\quad [\theta,\theta]_S= 0 \, .
\end{equation}

Both Dirac structures impose restriction on string fluxes. In order to construct Courant algebroid so that we can lift some of these restrictions on Dirac structures, we introduce twisted Courant brackets.

\section{Twisted Courant bracket}
\cleq




In this paper, we adopt an approach that allows for twisting of the Courant bracket by an arbitrary $O(D, D)$ transformation, i.e. the transformation that preserves the inner product $(\ref{eq:skalproizvod})$ 
\begin{equation} \label{eq:condort}
\langle {\cal O} \Lambda_1, {\cal O} \Lambda_2 \rangle = \langle  \Lambda_1, \Lambda_2 \rangle  \Leftrightarrow ({\cal O}\Lambda_1)^T\ \eta\ ({\cal O}\Lambda_2) = \Lambda^T_1\ \eta\ \Lambda_2 \, ,
\end{equation}
which is satisfied for the condition
\begin{equation} \label{eq:condorth}
{\cal O}^T\ \eta\ {\cal O} = \eta \, .
\end{equation}
The solutions for the above equation are in the form ${\cal O} = e^T$, where \cite{gualtieri}
\begin{equation} \label{eq:Tmatrix}
T = 
\begin{pmatrix}
A & \theta \\
B & -A^T 
\end{pmatrix} \, ,
\end{equation}
with $\theta : T^\star {\cal M} \to T {\cal M}$ and $B: T{\cal M} \to T^\star {\cal M}$ being antisymmetric, and $A: T{\cal M} \to T{\cal M}$, being the endomorphisms on the space of vectors. In general case, $B$ and $\theta$ can be independent for ${\cal O}$ to satisfy condition (\ref{eq:condorth}).  If ${\cal O} = e^T$, then it is the element of the $O(D,D)$ group \cite{gualtieri}. 

In order to see how twisted Courant brackets appear in the algebra of symmetry generators, we note that the relation (\ref{eq:GGcourant}) can be written as
\begin{equation} \label{eq:twdef1}
\int d\sigma_1 d\sigma_2 \Big\{ \langle \Lambda_1 , X\rangle (\sigma_1) , \langle\Lambda_2, X \rangle (\sigma_2)\Big\} = - \int d\sigma\langle[\Lambda_1 , \Lambda_2 ]_{\cal C} , X  \rangle (\sigma) \, .
\end{equation} 
Consider now the action of some element of $O(D,D)$ on the double coordinate $X$ (\ref{eq:Xdouble}) and the double gauge parameter $\Lambda$ (\ref{eq:Lxi})
\begin{equation} \label{eq:hatXL}
\hat{X}^M = {\cal O}^M_{\ N}\ X^N \, ,\ \hat{\Lambda}^M =  {\cal O}^M_{\ N}\ \Lambda^N \, ,
\end{equation}
Using (\ref{eq:condort}) and (\ref{eq:hatXL}), the relation (\ref{eq:twdef1}) can be written as 
\begin{eqnarray} \label{eq:twdef2}
\int d\sigma_1 d\sigma_2  \Big\{ \langle\hat{\Lambda}_1, \hat{X} \rangle (\sigma_1), \langle \hat{\Lambda}_2,  \hat{X} \rangle (\sigma_2) \Big\}&=& - \int d\sigma \langle [\Lambda_1, \Lambda_2]_{\cal C}, X \rangle  (\sigma) \\\notag
&=&-\int d\sigma \langle [\hat{\Lambda}_1, \hat{\Lambda}_2]_{{\cal C}_T}, \hat{X}\rangle  (\sigma) \, ,
\end{eqnarray}
where we expressed the right hand side in terms of some new bracket $[\hat{\Lambda}_1,\hat{\Lambda}_2]_{{\cal C}_T} $. 
The left hand side is the Poisson bracket of two inner product of two generalized vectors. It can be recognized as the Poisson bracket algebra of two generators of the form 
\begin{equation} \label{eq:GTdef}
{\cal \hat{G}}_{\hat{\Lambda}} = \int d\sigma \langle \hat{\Lambda} , \hat{X} \rangle \, .
\end{equation}

Moreover, using (\ref{eq:hatXL}) and (\ref{eq:condort}), the right hand side of (\ref{eq:twdef1}) can be written as
\begin{equation} \label{eq:twdef3}
\langle [\Lambda_1, \Lambda_2]_{\cal C}, X\rangle =\langle[{\cal O}^{-1}\hat{\Lambda}_1, {\cal O}^{-1}\hat{\Lambda}_2]_{\cal C}, {\cal O}^{-1}\hat{X}\rangle = \langle{\cal O} [{\cal O}^{-1}\hat{\Lambda}_1, {\cal O}^{-1}\hat{\Lambda}_2]_{\cal C}, \hat{X}\rangle \, .
\end{equation}
Using (\ref{eq:twdef1}), (\ref{eq:twdef2}) and (\ref{eq:twdef3}), one obtains
\begin{equation} \label{eq:twdef}
[\hat{\Lambda}_1,\hat{\Lambda}_2]_{{\cal C}_T} = {\cal O} [{\cal O}^{-1}\hat{\Lambda}_1, {\cal O}^{-1}\hat{\Lambda}_2]_{\cal C} = e^T [e^{-T}\hat{\Lambda}_1, e^{-T}\hat{\Lambda}_2]_{\cal C} \, .
\end{equation}
This is a definition of $T$-twisted Courant bracket. We see that it can be easily obtained from the Poisson bracket algebra of two generators written in a suitable basis, as in (\ref{eq:GTdef}).

Throughout this paper, we use the notation where the  $[,]_{\cal C}$ is the Courant bracket, while when ${\cal C}$ is accompanied with an index, it represents the twist of the Courant bracket by the indexed field, e.g. $[,]_{{\cal C}_B}$ is the Courant bracket twisted by $B$.

\subsection{Courant algebroid}

The relation (\ref{eq:twdef}) tells us that an element of the $O(D,D)$ group defines an isomorphism between different brackets
\begin{equation}
e^T[\hat{\Lambda}_1,\hat{\Lambda}_2]_{{\cal C}} = [e^T\hat{\Lambda}_1, e^T\hat{\Lambda}_2]_{{\cal C}_T} \, ,
\end{equation}
while keeping the inner product invariant. We want to see if this isomorphism between the brackets translate into the isomorphism between the Courant algebroids as well. In order to complete the construction of the Courant algebroids that can be defined with the twisted Courant brackets we are looking for an anchor that satisfies
\begin{equation}
\rho \Big( [\Lambda_1, \Lambda_2]_{\cal C_{T}} \Big) =  [\rho (\Lambda_1), \rho (\Lambda_2) ]_L \, .
\end{equation}
Using (\ref{eq:twdef}), we rewrite the previous relation as
\begin{equation}
\rho \Big(  e^T [e^{-T}\Lambda_1, e^{-T}\Lambda_2]_{\cal C} \Big) =  [\rho (\Lambda_1), \rho (\Lambda_2) ]_L = [\rho \Big( e^T (e^{-T} \Lambda_1) \Big), \rho \Big( e^T  (e^{-T} \Lambda_2) \Big) ]_L \, .
\end{equation}
Now we can use the fact that the natural projection is the anchor for the standard Courant algebroid, with Courant bracket as its anchor bracket. The anchor for the twisted Courant bracket is then given by
\begin{equation} \label{eq:RHO}
\rho(\Lambda)  = \pi (e^{-T} \Lambda) \, .
\end{equation}

Next, the differential operator can be obtained from 
\begin{equation}
\langle {\cal D}f, \Lambda \rangle ={\cal L}_{\rho (\Lambda)}\ f = {\cal L}_{\pi  (e^{-T} \Lambda)} f  = \langle d f, e^{-T} \Lambda \rangle = \langle e^T d f, \Lambda \rangle \, ,
\end{equation}
where in the second step the definition of $\rho$ is used, in the next step we the Courant algebroid property for the natural projection was used, and in the last step the fact that $e^{-T}$ is orthogonal.  The differential operator is given by
\begin{equation} \label{eq:DDdef}
{\cal D} f= e^{T} df \, .
\end{equation}

The anchor $\rho$ (\ref{eq:RHO}) was defined so that it satisfies the first Courant algebroid compatibility condition (\ref{eq:CAdef1}). Now we verify the remaining conditions.
We have
\begin{eqnarray}
[\Lambda_1, f \Lambda_2]_{\cal C_{T}} &=& e^{T}[e^{-T}\Lambda_1, f e^{-T}\Lambda_2]_{\cal C} \\ \notag
&=& e^{T} \Big(f [e^{-T} \Lambda_1, e^{-T} \Lambda_2]_{\cal C} + ({\cal L}_{\pi (e^{-T} \Lambda_1)} f) (e^{-T}\Lambda_2) - \frac{1}{2} \langle e^{-T} \Lambda_1, e^{-T} \Lambda_2 \rangle  df \Big) \\ \notag
&=& f [\Lambda_1, \Lambda_2]_{\cal C_{T}} + ({\cal L}_{\rho (\Lambda_1)} f) \Lambda_2 - \frac{1}{2} \langle \Lambda_1, \Lambda_2 \rangle {\cal D}f \, ,
\end{eqnarray}
where we firstly used definition of the twisted Courant bracket (\ref{eq:twdef}), afterwards we applied the second compatibility condition (\ref{eq:CAdef2}) for the Courant bracket, and in the end used the expressions for the anchor $\rho$ (\ref{eq:RHO}) and the differential operator ${\cal D}$ (\ref{eq:DDdef}), as well as the fact that $O(D,D)$ transformations keep the inner product invariant (\ref{eq:condort}).

For the following condition (\ref{eq:CAdef3}), we firstly write
\begin{eqnarray}
\langle [\Lambda_1, \Lambda_2]_{\cal C_T} +\frac{1}{2} {\cal D}\langle \Lambda_1, \Lambda_2 \rangle , \Lambda_3 \rangle &=& \langle e^{T} [e^{-T} \Lambda_1, e^{-T} \Lambda_2]_{\cal C} +\frac{1}{2} e^{T} d \langle \Lambda_1,  \Lambda_2 \rangle , \Lambda_3 \rangle \\ \notag
&=& \langle [e^{-T} \Lambda_1, e^{-T} \Lambda_2]_{\cal C} +\frac{1}{2}  d \langle e^{-T} \Lambda_1,   e^{-T} \Lambda_2 \rangle , e^{-T} \Lambda_3 \rangle  \, ,
\end{eqnarray}
and
\begin{equation}
{\cal L}_{\rho(\Lambda_1)} \langle \Lambda_2, \Lambda_3 \rangle = {\cal L}_{\pi (e^{-T} \Lambda_1)} \langle e^{-T} \Lambda_2, e^{-T} \Lambda_3 \rangle \, .
\end{equation}
From the previous two relations it is obvious that the third compatibility condition is equivalent to the respective condition for the standard Courant algebroid for generalized vectors $e^{-T} \Lambda_i \, , \  i \in 1,\ 2, \ 3$. 

The fourth condition (\ref{eq:CAdef4}) is as easily obtained from the orthogonality of $e^T$
\begin{equation}
\langle {\cal D} f, {\cal D} g \rangle  = \langle e^T d f, e^T d g \rangle  = \langle d f,d g \rangle = 0 \, .
\end{equation}

Lastly, we note that
\begin{equation}
[[\Lambda_1, \Lambda_2]_{\cal C_T}, \Lambda_3 ]_{\cal C_T} = e^{T} [[e^{-T}  \Lambda_1, e^{-T}  \Lambda_2]_{\cal C}, e^{-T} \Lambda_3 ]_{\cal C} \, ,
\end{equation}
from which we write the Jacobiator for the twisted Courant bracket 
\begin{equation} \label{eq:JacT}
\text{Jac}_{{\cal C}_{T}}(\Lambda_1,\Lambda_2,\Lambda_3) =e^{T}\text{Jac}_{\cal C}(e^{-T}\Lambda_1,e^{-T}\Lambda_2,e^{-T}\Lambda_3)\, .
\end{equation}
Similarly, we note that
\begin{equation}
\langle [\Lambda_1, \Lambda_2]_{\cal C_T}, \Lambda_3 \rangle = \langle e^{T} [e^{-T} \Lambda_1, e^{-T} \Lambda_2]_{\cal C}, \Lambda_3 \rangle = \langle  [e^{-T} \Lambda_1, e^{-T} \Lambda_2]_{\cal C}, e^{-T} \Lambda_3 \rangle \, ,
\end{equation}
from which one easily obtains the Nijenhuis operator 
\begin{equation}\label{eq:NijT}
\text{Nij}_{{\cal C}_T} ( \Lambda_1,\Lambda_2,\Lambda_3) =\text{Nij}_{\cal C} (e^{-T}\Lambda_1,e^{-T}\Lambda_2,e^{-T}\Lambda_3)\, .
\end{equation}
Using (\ref{eq:DDdef}), (\ref{eq:JacT}) and (\ref{eq:NijT}), we obtain the last compatibility condition (\ref{eq:CAdef5}).

We have proven that each $O(D, D)$ transformation defines a Courant algebroid $\Big(T{\cal M}\oplus T^\star{\cal M} ,\langle, \rangle, [,]_{{\cal C}_T}, \rho \Big)$ with the appropriate twisted Courant bracket and the natural inner product (\ref{eq:skalproizvod}). As we consider all $O(D,D)$ transformations, including those that are not automorphisms of the Courant bracket, we can derive various twisted Courant brackets along with their corresponding Courant algebroids and Dirac structures, as will be discussed in the following chapters.
\section{B-twisted Courant bracket}
\cleq

The simplest case and the first one to consider is the one of the $B$-twisted Courant bracket. In mathematical literature, this bracket was firstly obtained in \cite{courant1}, by adding the 3-form term contracted with the vector components to the expression for the standard Courant bracket. 

The $B$-twisted Courant bracket is defined by
\begin{equation} \label{eq:CourantB}
[\Lambda_1, \Lambda_2 ]_{{\cal C}_B} = e^{\hat{B}} [e^{-\hat{B}} \Lambda_1, e^{-\hat{B}}\Lambda_2 ]_{\cal C} \, ,
\end{equation}
where the $O(D,D)$ transformation governing twist by a 2-form $B$ is obtained by substituting $A= 0$, $\theta = 0$ in (\ref{eq:Tmatrix}), which results in 
\begin{equation} \label{eq:ebhat}
e^{\hat{B}} = \begin{pmatrix}
\delta^\mu_\nu & 0 \\
2B_{\mu \nu} & \delta^\nu_\mu
\end{pmatrix}\, , \ \ 
\hat{B}^M_{\ N} = 
\begin{pmatrix}
0 & 0 \\
2B_{\mu \nu} & 0 \\ 
\end{pmatrix}\, .
\end{equation}
In \cite{cdual}, the $B$-twisted Courant bracket was obtained from
\begin{equation}
\Big\{ {\cal G}^B_{\Lambda_1},  {\cal G}^B_{\Lambda_2} \Big \} = - {\cal G}^B_{[\Lambda_1, \Lambda_2]_{{\cal C}_B}}  \, ,
\end{equation}
where ${\cal G}^B$ is the generator, given by
\begin{equation}
{\cal G}^B_{\Lambda}= \int d\sigma \Big( \xi^\mu i_\mu + \lambda_\mu \kappa x^{\prime \mu} \Big) \, , 
\end{equation}
and $i_\mu$ are basis currents, given by
\begin{equation} \label{eq:idef}
i_\mu = \pi_\mu + 2\kappa B_{\mu \nu} x^{\prime \nu} \, .
\end{equation}
The basis currents give rise to the $H$-flux in their Poisson bracket algebra
\begin{equation} \label{eq:ii}
\{ i_\mu (\sigma), i_\nu (\bar{\sigma}) \} = -2 \kappa B_{\mu \nu \rho} x^{\prime \rho} \delta(\sigma- \bar{\sigma}) \, ,
\end{equation}
where $B_{\mu \nu \rho}$ is the Kalb-Ramond field strength (\ref{eq:bmnr}).

The bracket (\ref{eq:CourantB}) results in a generalized vector $\Lambda = \xi \oplus \lambda$, where \cite{cdual}
\begin{equation} \label{eq:XIB}
\xi^\mu = \xi_1^\nu \partial_\nu \xi_2^\mu - \xi_2^\nu \partial_\nu \xi_1^\mu \, ,
\end{equation} and
\begin{equation} \label{eq:LB}
\lambda_\mu = \xi_1^\nu (\partial_\nu \lambda_{2 \mu} - \partial_\mu \lambda_{2 \nu}) - \xi_2^\nu (\partial_\nu \lambda_{1 \mu} - \partial_\mu \lambda_{1 \nu}) +\frac{1}{2} \partial_\mu (\xi_1 \lambda_2- \xi_2 \lambda_1 )+ 2  B_{\mu \nu \rho} \xi^\nu_1 \xi^\rho_2 \, ,
\end{equation}
or in the coordinate free notation
\begin{eqnarray}
\xi &=& [\xi_1, \xi_2]_L \, ,\\ \notag
\lambda &=& {\cal L}_{\xi_1}\lambda_2 -{\cal L}_{\xi_2}\lambda_1 - \frac{1}{2}d (i_{\xi_1}\lambda_2-i_{\xi_2}\lambda_1) + i_{\xi_1} i_{\xi_2} d B \, .
\end{eqnarray}

The $B$-twisted Courant bracket, together with the $O(D,D)$ invariant inner product (\ref{eq:skalproizvod}) defines a Courant algebroid. Using (\ref{eq:ebhat}) and (\ref{eq:RHO}), we obtain the anchor
\begin{equation} \label{eq:rhoB}
\rho_B  = \pi e^{-\hat{B}} = \begin{pmatrix}
1 & 0 \\
0 & 0 
\end{pmatrix}  \begin{pmatrix}
1 & 0 \\
-2B & 1
\end{pmatrix}
=
\begin{pmatrix}
1 & 0 \\
0 & 0 
\end{pmatrix}
= \pi \, .
\end{equation}
Similarly, substituting (\ref{eq:ebhat}) into (\ref{eq:DDdef}), we obtain the differential operator 
\begin{equation}
D_B f = \begin{pmatrix}
1 & 0 \\
2B & 1
\end{pmatrix} \begin{pmatrix}
0 \\
d f
\end{pmatrix} =  \begin{pmatrix}
0 \\
d f
\end{pmatrix} \, .
\end{equation}
Both the anchor and the differential operator are the same as in case of the standard Courant algebroid.

\subsection{Dirac structures}

Primarily, we consider the subspace of parameters ${\cal I}_B$ in the form of (\ref{eq:isoB}). Then, the generator is given by
\begin{equation}
{\cal G}^B_{{\cal I}_B (\Lambda)} = \int d\sigma \Big(\xi^\mu i_\mu +  2 B_{\mu \nu}\xi^\nu \kappa x^{\prime \mu}   \Big) \, .
\end{equation}
After substituting (\ref{eq:idef}) into the previous relation and taking into the account the antisymmetry of $B$, we obtain
\begin{equation}
{\cal G}^B_{{\cal I}_B(\Lambda)} = \int d\sigma \xi^\mu \pi_\mu  \, .
\end{equation}
This is just the diffeomorphism generator, which gives rise to the Lie bracket
\begin{equation}
\Big \{ {\cal G}^B_{{\cal I}_B(\Lambda_1)} ,  {\cal G}^B_{{\cal I}_B(\Lambda_2)}\Big \} = - {\cal G}^B_{[\xi_1, \xi_2]_L} \, .
\end{equation}
The sub-bundles are Dirac structures if and only if the bracket satisfies the Jacobi identity on them. Therefore, since the $B$-twisted Courant bracket reduces to the Lie bracket, the subspace ${\cal I}_B$ is Dirac structure with no additional constraints on $B$ imposed. Therefore, in the theory with non zero $H$-flux, the $B$-twisted Courant bracket has Dirac structures in the form of graph of $B$ over the tangent bundle
\begin{equation}
\Big[ {\cal I}_B(\Lambda_1), {\cal I}_B(\Lambda_2) \Big]_{{\cal C}_B} = {\cal I}_B \Big( [\Lambda_1,\Lambda_2]_{{\cal C}_B} \Big) \, .
\end{equation}
Secondarily, we consider the subspace ${\cal I}_{\theta}$ in the form (\ref{eq:isoth}). We have
\begin{equation}
{\cal G}^B_{{\cal I}_{\theta} (\Lambda) } = \int d\sigma \Big(\kappa \theta^{\mu \nu} \lambda_{\nu} i_\mu +  \lambda_\mu \kappa  x^{\prime \mu}   \Big) \, ,
\end{equation}
The simplest way to obtain the conditions for Dirac structure is to substitute $\xi = \kappa \theta^{\mu \nu} \lambda_\nu$ into (\ref{eq:XIB}) and (\ref{eq:LB}). Similarly as in one of the previous chapters, we obtain 
\begin{equation}
\lambda_{\mu} = \theta^{\nu \rho} (\lambda_{1\rho} \partial_{\nu} \lambda_{2\mu}-\lambda_{2\rho} \partial_{\nu} \lambda_{1\mu}) + \partial_\mu \theta^{\nu \rho} \lambda_{1\rho} \lambda_{2\nu} + 2 \kappa^2  B_{\mu \nu \rho} \theta^{\nu \alpha} \theta^{\rho \beta} \lambda_{1\alpha} \lambda_{2 \beta} \, ,
\end{equation}
while the vector component is given by
\begin{eqnarray}
\xi^\mu &=& \kappa^2 \theta^{\mu \sigma} \theta^{\nu \rho} (\lambda_{1\rho}\partial_\nu \lambda_{2\sigma}-\lambda_{2\rho}\partial_\nu \lambda_{1\sigma}) + (\theta^{\nu \rho}\partial_\nu \theta^{\mu \sigma}-\theta^{\nu \sigma}\partial_\nu \theta^{\mu \rho})\lambda_{1\rho}\lambda_{2\sigma} \\ \notag
& =& \theta^{\mu \sigma} \lambda_{\sigma} +R^{\mu \nu \rho } \lambda_{1\nu} \lambda_{2\rho} - 2\kappa^3 \theta^{\mu \alpha} \theta^{\nu \beta} \theta^{\rho \gamma} B_{\alpha \beta \gamma} \lambda_{1\nu} \lambda_{2\rho} \, .
\end{eqnarray}
We conclude that the condition for ${\cal I}_{\theta}$ to be a Dirac structure is given by
\begin{equation} \label{eq:Rcal}
{\cal R}^{\mu \nu \rho}  = R^{\mu \nu \rho } - 2 \kappa^3 \theta^{\mu \alpha} \theta^{\nu \beta} \theta^{\rho \gamma} B_{\alpha \beta \gamma } = 0\, ,
\end{equation}
where ${\cal R}^{\mu \nu \rho}$ is the generalized ${\cal R}$-flux. The obtained condition defines WZW-Poisson structures \cite{WZW}. 

\section{$\theta$-twisted Courant bracket}
\cleq

The next case that we will consider is the $\theta$-twisted Courant bracket. This bracket was obtained in the algebra of currents T-dual to the currents that give rise to the $B$-twisted Courant bracket \cite{crdual}. The twist is governed by the $\theta$-transformations, obtained by substituting  $A=0$ and $B=0$ into (\ref{eq:Tmatrix}), resulting in 
\begin{equation} \label{eq:enateta}
e^{\hat{\theta}} = 
\begin{pmatrix}
\delta^\mu_\nu & \kappa \theta^{\mu \nu} \\
0 & \delta^\nu_\mu
\end{pmatrix} \, , \ \ 
\hat{ \theta}^M_{\ N} = 
\begin{pmatrix}
0 & \kappa \theta^{\mu \nu} \\
0 & 0 
\end{pmatrix} \, .
\end{equation}
The bracket (\ref{eq:twdef}) becomes the Courant bracket twisted by a bi-vector $\theta$
\begin{equation} \label{eq:CourantTheta}
[\Lambda_1, \Lambda_2 ]_{{\cal C}_\theta} = e^{\hat{\theta}} [e^{-\hat{\theta}} \Lambda_1,e^{-\hat{\theta}} \Lambda_2 ]_{\cal C} \, ,
\end{equation}
which is the same bracket as the one obtained in the Poisson bracket algebra of the generator \cite{cdual}
\begin{equation} \label{eq:GGGth}
\{{\cal G}^{\theta}_{\Lambda_1},  {\cal G}^{\theta}_{\Lambda_2}\} =- {\cal G}^{\theta}_{\Lambda} \, , \quad {\cal G}^{\theta}_{\Lambda} = \int d \sigma \Big( \xi^\mu \pi_\mu + \lambda_\mu k^\mu \Big) \, ,
\end{equation}
where $k^\mu$ are currents, given by
\begin{equation}\label{eq:kdef}
k^\mu = \kappa x^{\prime \mu} + \kappa \theta^{\nu \rho} \pi_\rho \, .
\end{equation}
These currents give rise to the non-geometric fluxes in its algebra by
\begin{equation}
\{ k^\mu, k^\nu \} = -\kappa Q_\rho^{\ \mu \nu} k^\rho - \kappa^2 R^{\mu \nu \rho} \pi_\rho \, ,
\end{equation}
where
\begin{equation} \label{eq:nongeomflux1}
Q_\mu^{\ \nu \rho} = \partial_\mu \theta^{\nu \rho} \, .
\end{equation}

The full expression for the vector and 1-form components that result from the bracket is given by
\begin{align} \label{eq:XIR}
\xi^\mu =&\ \xi_1^\nu \partial_\nu \xi_2^\mu - \xi_2^\nu \partial_\nu\xi_1^\mu + \\ \notag
& +\kappa \theta^{\mu \nu}\Big( \xi_1^\rho (\partial_\rho \lambda_{2 \nu}-\partial_\nu \lambda_{2 \rho}) - \xi_2^\rho (\partial_\rho \lambda_{1 \nu}- \partial_\nu \lambda_{1 \rho}) +\frac{1}{2} \partial_\nu (\xi_1 \lambda_{2} - \xi_2 \lambda_1) \Big) \\ \notag
& + \kappa \xi_1^\nu \partial_\nu (\lambda_{2 \rho} \theta^{\rho \mu})-\kappa \xi_2^\nu \partial_\nu (\lambda_{1 \rho} \theta^{\rho \mu})+\kappa (\lambda_{1 \nu} \theta^{\nu \rho}) \partial_\rho \xi_2^\mu -\kappa (\lambda_{2 \nu}\theta^{\nu \rho}) \partial_\rho \xi_1^\mu \\ \notag
&+\kappa^2 R^{\mu \nu \rho} \lambda_{1 \nu}\lambda_{2 \rho} \, ,
\end{align}
and
\begin{align} \label{eq:LR}
\lambda_\mu = &\ \xi_1^\nu (\partial_\nu \lambda_{2 \mu} - \partial_\mu \lambda_{2 \nu}) - \xi_2^\nu (\partial_\nu \lambda_{1 \mu} - \partial_\mu \lambda_{1 \nu}) +\frac{1}{2}\partial_\mu(\xi_1 \lambda_2 - \xi_2 \lambda_1) \\ \notag
& + \kappa \theta^{\nu \rho} (\lambda_{1 \nu}\partial_\rho \lambda_{2 \mu}-\lambda_{2 \nu} \partial_\rho \lambda_{1 \mu})+ \kappa \lambda_{1 \rho} \lambda_{2 \nu} Q_\mu^{\ \rho \nu} \, .
\end{align}
In the coordinate free notation, the relations above become 
\begin{eqnarray}
\xi &=& [\xi_1, \xi_2]_L +\kappa \theta  \Big( {\cal L}_{\xi_1}\lambda_2-{\cal L}_{\xi_2}\lambda_1 - \frac{1}{2}d (i_{\xi_1}\lambda_2 -i_{\xi_2}\lambda_1  ) \Big)\, , \\ \notag
&& + [\xi_1, \kappa \lambda_2 \theta]_L-[\xi_2, \kappa \lambda_1 \theta]_L+ \frac{\kappa^2}{2} i _{\lambda_1} i_{\lambda_2} [\theta, \theta]_S \\ \notag
\lambda &=& {\cal L}_{\xi_1}\lambda_2-{\cal L}_{\xi_2}\lambda_1 - \frac{1}{2}d (i_{\xi_1}\lambda_2 -i_{\xi_2}\lambda_1 )+\kappa  [\lambda_1, \lambda_2]_{\theta} \, ,
\end{eqnarray}
where $(\lambda \theta)^\mu = \lambda_\nu \theta^{\nu \mu}$ and $(\theta \lambda)^\mu = \theta^{\mu \nu} \lambda_\nu$.

As per our analysis in the third chapter, this bracket defines a Courant algebroid. Its anchor is obtained from (\ref{eq:enateta}) and (\ref{eq:RHO})
\begin{equation}
\rho_{\theta}  = \pi e^{-\hat{\theta}} = \begin{pmatrix}
1 & 0 \\
0 & 0 
\end{pmatrix} \begin{pmatrix}
1 & -\kappa \theta \\
0 & 1 
\end{pmatrix} = 
\begin{pmatrix}
1 & -\kappa \theta \\
0 & 0 
\end{pmatrix}  \, .
\end{equation}
The differential operator is obtained from (\ref{eq:DDdef}), and is given by
\begin{equation}
{\cal D}_{\theta} f = \begin{pmatrix}
1 & \kappa \theta \\
0 & 1
\end{pmatrix}
\begin{pmatrix}
0 \\ 
d f
\end{pmatrix} = 
\begin{pmatrix}
\kappa \theta d f \\
d f
\end{pmatrix}
\, .
\end{equation}
The differential operator can also be written as 
\begin{equation}
{\cal D}_{\theta} f = \begin{pmatrix}
d_\theta f \\
d f
\end{pmatrix} \, , 
\end{equation}
where $d_\theta f$ is defined in (\ref{eq:ext-der-theta}). 

\subsection{Dirac structures}

Now consider the subspace ${\cal I}_B$ (\ref{eq:isoB}).  If we substitute $\lambda_\mu = 2 B_{\mu \nu} \xi^\nu$ and (\ref{eq:kdef}) into (\ref{eq:GGGth}), we obtain
\begin{eqnarray} \label{eq:Gth1}
{\cal G}^{\theta}_{{\cal I}_{B}(\Lambda) }&=& \int d\sigma \Big( \xi^\mu \pi_\mu + 2 B_{\mu \nu} \xi^\nu \kappa \theta^{\mu \rho} \pi_\rho + 2 B_{\mu \nu} \kappa x^{\prime \mu} \xi^\nu \Big)  \\ \notag
&=& \int d\sigma \Big( \xi^\nu (\delta^\mu_\nu - 2 \kappa B_{\nu \rho} \theta^{\rho \mu})\pi_\mu + 2B_{\mu \nu } \kappa x^{\prime \nu}\xi^\nu \Big)  \\ \notag
&=& \int d\sigma \Big( \tilde{\xi}^\mu \pi_\mu + \tilde{\lambda}_\mu \kappa x^{\prime \mu} \Big) \, ,
\end{eqnarray}
where 
\begin{equation}
\tilde{\xi}^\mu = \xi^\mu - 2 \kappa \theta^{\mu \nu} B_{\nu \rho} \xi^\rho \, , \ \quad \tilde{\lambda}_\mu = 2B_{\mu \nu} \xi^\nu \, .
\end{equation}
We can further simplify this using the following decomposition of the identity
\begin{equation} \label{eq:decomp}
\delta^\mu_\nu = 2\kappa \theta^{\mu \rho} B_{\rho \nu} + (G_E^{-1})^{\mu \rho} G_{\rho \nu} \, ,
\end{equation}
which can be easily obtained from (\ref{eq:thetadef}) and (\ref{eq:Geff}). This allows us to write the relation between the new parameters as
\begin{equation} \label{eq:tildecond}
\tilde{\lambda}_\mu = 2 (B G^{-1} G_E )_{\mu \nu} \tilde{\xi}^\nu
\end{equation}
The algebra of generator (\ref{eq:Gth1}) gives rise to the Courant bracket. Using the condition (\ref{eq:tildecond}) and (\ref{eq:IBIB}), we see that the sub-bundle ${\cal I}_B$ defines a Dirac structure if 
\begin{equation}
d (B G^{-1} G_E) = d B - 4 d (B^3) = 0  \, .
\end{equation}
where the cube of Kalb-Ramond field is defined by
\begin{equation} \label{eq:B-cube}
B^3 = B G^{-1} B G^{-1} B \, .
\end{equation}
As we see, the twisting by the bi-vector $\theta$ imposes condition on $H$ flux in order for ${\cal I}_B(\Lambda)$ to be a Dirac structure.

Let us now consider the subspace ${\cal I}_{\theta}$ (\ref{eq:isoth}). The generator on this subspace is given by
\begin{equation}
{\cal G}^{\theta}_{{\cal I}_{\theta}(\Lambda) }= \int d\sigma \Big(\kappa \theta^{\mu \nu} \lambda_{\nu} \pi_\mu + k^\mu \lambda_\mu \Big) = \int d\sigma  \kappa x^{\prime \mu} \lambda_\mu \, ,
\end{equation}
where we have used (\ref{eq:kdef}). This is the generator that does not depend on momenta $\pi_\mu$.  Hence, ${\cal I}_{\theta}$ are Dirac structures for the $\theta$-twisted Courant bracket, regardless of a bi-vector $\theta$. Moreover, we have a more general equation
\begin{equation} \label{eq:dthTh}
\Big[ {\cal I}_{\theta}(\Lambda_1), {\cal I}_{\theta}(\Lambda_2) \Big]_{{\cal C}_{\theta}} = 0 \, .
\end{equation}
We saw that the introduction of the non-commutative parameter removes the restriction from bi-vector for its graph to be a Dirac structure. However, there is a restriction on a 2-form now. This restriction is not independent of $\theta$. 

\section{Courant bracket twisted by $B$ and $\theta$}
\cleq

The next step in our analysis is considering introducing both twist by $B$ and by $\theta$. Due to $O(D,D)$ being a group, it is clear that the transformation given by
\begin{equation}
e^{\hat{\theta}} e^{\hat{B}} = \begin{pmatrix}
1 + 2\kappa \theta B & \kappa \theta \\
2 B & 1 
\end{pmatrix}
\end{equation}
remains in $O(D,D)$. It defines the Courant bracket that was firstly twisted by $B$ and then that bracket was twisted by $\theta$
\begin{eqnarray}\label{eq:CourantBTheta}
[\Lambda_1,  \Lambda_2]_{{\cal C}_{B, \theta}} =  e^{\hat{\theta}} e^{\hat{B}} [e^{-\hat{B}}e^{-\hat{\theta}}\Lambda_1, e^{-\hat{B}}e^{-\hat{\theta}} \Lambda_2]_{\cal C} = e^{\hat{\theta}} [e^{-\hat{\theta}}\Lambda_1, e^{-\hat{\theta}} \Lambda_2]_{{\cal C}_B} \, , 
\end{eqnarray}
where in the second step we applied the inverse relation of (\ref{eq:CourantB}). 
This bracket can be obtained from the Poisson bracket algebra of the generator 
\begin{equation} \label{eq:GBth}
{\cal G}^{B, \theta}_{\Lambda}= \int d\sigma \langle \Lambda, \hat{X} \rangle \, , 
\end{equation}
where
\begin{equation}
\hat{X} = e^{\hat{\theta}} e^{\hat{B}}  X = \begin{pmatrix}
\kappa x^{\prime \mu} + 2\kappa^2 (\theta B)^\mu_{\ \nu} x^{\prime \nu} + \kappa \theta^{\mu \nu} \pi_\nu \\
\pi_\mu + 2\kappa B_{\mu \nu} x^{\prime \nu}
\end{pmatrix} = 
\begin{pmatrix}
\kappa x^{\prime \mu} + \kappa \theta^{\mu \nu} i_\nu \\
i_\mu
\end{pmatrix} = 
\begin{pmatrix}
\hat{k}^\mu \\
i_\mu
\end{pmatrix} \, ,
\end{equation}
where $i_\mu$ is defined in  (\ref{eq:idef}). The algebra of currents $i_\mu$ and $\hat{k}^\mu$ was obtained in \cite{ nick1}. We have
\begin{eqnarray} \label{eq:ii-k}
\{ i_\mu (\sigma) , i_\nu (\bar{\sigma}) \} = -2 B_{\mu \nu \rho} \hat{k}^\rho \delta(\sigma - \bar{\sigma}) - {\cal F}_{\mu \nu}^{\ \rho}\ i_\rho \delta(\sigma-\bar{\sigma}) \, ,
\end{eqnarray}
where ${\cal F}$ is generalized $f$-flux, given by
\begin{equation} \label{eq:calF}
{\cal F}_{\mu \nu}^{\ \rho} = - 2\kappa B_{\mu \nu \sigma} \theta^{\sigma  \rho} \, .
\end{equation}
Note that (\ref{eq:ii-k}) is the same as the relation (\ref{eq:ii}), just expressed in terms of new basis currents $i_\mu$ and $\hat{k}^\mu$.
The other currents algebra is given by
\begin{eqnarray} \label{eq:khatkhat}
\{ \hat{k}^\mu (\sigma), \hat{k}^\nu (\bar{\sigma}) \} = - \kappa{\cal Q}_{\rho}^{\ \mu \nu} \hat{k}^\rho \delta(\sigma-\bar{\sigma}) -\kappa^2{\cal R}^{\mu \nu \rho} i_\rho \delta(\sigma-\bar{\sigma}) \, ,
\end{eqnarray}
where the generalized non-geometric $Q$ flux is given by
\begin{equation} \label{eq:calQ}
{\cal Q}^{\ \nu \rho}_\mu
 =   Q^{\ \nu \rho}_\mu + 2\kappa \theta^{\nu\sigma} \theta^{\rho \tau} B_{\mu \sigma \tau} \, , 
\end{equation}
and the generalized non-geometric $R$ flux by
\begin{equation} \label{eq:calR}
{\cal R}^{\mu \nu \rho} = R^{\mu \nu \rho} +2\kappa \theta^{\mu \lambda} \theta^{\nu \sigma}\theta^{\rho \tau}B_{\lambda \sigma \tau} \, .
\end{equation}
Lastly, we have
\begin{equation} \label{eq:kringiota}
\{ i_\mu (\sigma), \hat{k}^\nu (\bar{\sigma}) \} = \kappa \delta_\mu^\nu \delta^\prime(\sigma-\bar{\sigma}) + {\cal F}^{\ \nu}_{\mu \rho}\ \hat{k}^\rho \delta(\sigma-\bar{\sigma}) -\kappa {\cal Q}_{\mu}^{\ \nu \rho} i_\rho \delta(\sigma-\bar{\sigma}) \, .
\end{equation}

The full bracket (\ref{eq:CourantBTheta}) can then be obtained from the Poisson bracket algebra 
\begin{equation} \label{eq:GbthGbth}
\{ {\cal G}^{B, \theta}_{\Lambda_1}, {\cal G}^{B, \theta}_{\Lambda_2} \} = - {\cal G}^{B, \theta}_{\Lambda} \, , \qquad \Lambda = [\Lambda_1, \Lambda_2]_{{\cal C}_{B, \theta}} \, .
\end{equation}
The anchor for the Courant algebroid is given by
\begin{equation}
\rho_{B, \theta} =  \pi (e^{\hat{\theta}} e^{\hat{B}} )^{-1} = 
\begin{pmatrix}
1 & 0 \\
0 & 0 
\end{pmatrix}
 \begin{pmatrix}
1 &0 \\
-2B & 1 
\end{pmatrix} \begin{pmatrix}
1 & -\kappa \theta \\
0 & 1 
\end{pmatrix}= 
\begin{pmatrix}
1 & -\kappa \theta \\
0 & 0 
\end{pmatrix}  \, .
\end{equation}
The differential operator is given by
\begin{equation}
{\cal D}_{B, \theta} f= \begin{pmatrix}
1 + 2\kappa \theta B & \kappa \theta \\
2 B & 1 
\end{pmatrix}
\begin{pmatrix}
0 \\ 
d f
\end{pmatrix} = 
\begin{pmatrix}
\kappa \theta d f \\
d f
\end{pmatrix}
\, .
\end{equation}
The full bracket was obtained in \cite{nick1}, from the algebra of charges
\begin{equation}
{\cal Q}_{R}(\xi, \lambda) = \int d\sigma \Big( \xi^\mu \pi_\mu + (\kappa x^{\prime \mu} + \kappa \theta^{\mu \nu} \pi_\nu) \lambda_\mu \Big) \, ,
\end{equation}
where the phase space was changed so that $\pi_\mu \to \pi_\mu + 2\kappa B_{\mu \nu} x^{\prime \nu}$. This effectively results in the generator (\ref{eq:GBth}), which gives the Courant bracket twisted firstly by $B$ and then by $\theta$  (\ref{eq:GbthGbth}), which results in a generalized vector with the vector component
\begin{align} \label{eq:XIRR}
\xi^\mu =&\ \xi_1^\nu \partial_\nu \xi_2^\mu - \xi_2^\nu \partial_\nu\xi_1^\mu + \\ \notag
& +\kappa \theta^{\mu \nu}\Big( \xi_1^\rho (\partial_\rho \lambda_{2 \nu}-\partial_\nu \lambda_{2 \rho}) - \xi_2^\rho ( \partial_\rho \lambda_{1 \nu}-\partial_\nu \lambda_{1 \rho}) +\frac{1}{2} \partial_\nu (\xi_1 \lambda_{2} - \xi_2 \lambda_1) \Big) \\ \notag
& + \kappa \xi_1^\nu \partial_\nu (\lambda_{2 \rho} \theta^{\rho \mu})-\kappa \xi_2^\nu \partial_\nu (\lambda_{1 \rho} \theta^{\rho \mu})+\kappa (\lambda_{1 \nu} \theta^{\nu \rho}) \partial_\rho \xi_2^\mu -\kappa (\lambda_{2 \nu}\theta^{\nu \rho}) \partial_\rho \xi_1^\mu +\kappa^2 R^{\mu \nu \rho} \lambda_{1 \nu}\lambda_{2 \rho}\\ \notag
&+ 2\kappa \theta^{\mu \tau}B_{\tau \rho \sigma}\xi_1^{\rho} \xi_2^{\sigma}+ 2\kappa^2 \theta^{\mu \sigma} \theta^{\nu \tau} B_{\tau \sigma \rho} (\xi_1^\rho \lambda_{2\nu}-\xi_2^\rho \lambda_{1\nu}) + 2\kappa^3 \theta^{\mu \sigma} \theta^{\nu \tau} \theta^{\rho \lambda} B_{\lambda \sigma \tau} \lambda_{1\nu} \lambda_{2\rho} \, ,
\end{align}
and 1-form component 
\begin{align} \label{eq:LRR}
\lambda_\mu = &\ \xi_1^\nu (\partial_\nu \lambda_{2 \mu} - \partial_\mu \lambda_{2 \nu}) - \xi_2^\nu (\partial_\nu \lambda_{1 \mu} - \partial_\mu \lambda_{1 \nu}) +\frac{1}{2}\partial_\mu(\xi_1 \lambda_2 - \xi_2 \lambda_1) \\ \notag
& + \kappa \theta^{\nu \rho} (\lambda_{1 \nu}\partial_\rho \lambda_{2 \mu}-\lambda_{2 \nu} \partial_\rho \lambda_{1 \mu})+ \kappa \lambda_{1 \rho} \lambda_{2 \nu} Q_\mu^{\ \rho \nu} \\ \notag
& 2 B_{\mu \nu \rho} \xi^{\nu}_1 \xi^\rho_2 + 2\kappa \theta^{\nu \rho} B_{\rho \mu \sigma} (\xi_1^\sigma \lambda_{2\nu} - \xi_2^\sigma \lambda_{1\nu} ) + 2\kappa^2 \theta^{\nu \sigma} \theta^{\rho \tau} B_{\tau \mu \sigma} \lambda_{1\nu} \lambda_{2\rho} \, .
\end{align}
In the coordinate free notation, the above relations become
\begin{eqnarray} 
\xi &=& [\xi_1, \xi_2]_L +\kappa\theta \Big( {\cal L}_{\xi_1}\lambda_2 -{\cal L}_{\xi_2}\lambda_1 - \frac{1}{2}d(i_{\xi_1}\lambda_2-i_{\xi_2}\lambda_1) \Big)  \\ \notag
&&+ [\xi_1, \kappa \lambda_2 \theta]_L- [\xi_2, \kappa \lambda_1 \theta]_L + \frac{\kappa^2}{2} i_{\lambda_1} i_{\lambda_2} [\theta, \theta]_S \\ \notag
&&+ \kappa\theta\Big(i_{\xi_1} i_{\xi_2} d B + i_{\xi_1} i_{\kappa \lambda_2 \theta}\ dB - i_{\xi_2} i_{\kappa \lambda_1 \theta}\ dB + i_{\kappa \lambda_1 \theta}\ i_{\kappa \lambda_2  \theta}\ d B \Big)   \, ,\\ \notag
\lambda &=& {\cal L}_{\xi_1}\lambda_2 -{\cal L}_{\xi_2}\lambda_1 - \frac{1}{2}d(i_{\xi_1}\lambda_2-i_{\xi_2}\lambda_1) +\kappa [\lambda_1, \lambda_2]_{\theta}  \\ \notag
&& i_{\xi_1} i_{\xi_2} d B + i_{\xi_1} i_{\kappa \lambda_2 \theta}\ dB - i_{\xi_2} i_{\kappa \lambda_1 \theta}\ dB + i_{\kappa \lambda_1 \theta}\ i_{\kappa \lambda_2  \theta}\ d B \, .
\end{eqnarray}

\subsection{Dirac structures}

On the sub-bundle ${\cal I}_B$, the generator (\ref{eq:GBth}) becomes
\begin{eqnarray}
{\cal G}^{B, \theta}_{{\cal I}_{B}(\Lambda)}  &=& \int d\sigma (\xi^\mu i_\mu + 2 \hat{k}^\mu B_{\mu \nu} \xi^\nu ) = \int d\sigma \xi^\mu (i_\mu - 2 B_{\mu \nu} \hat{k}^\nu) \\ \notag
&=& \int d\sigma \xi^\mu (\pi_\mu  - 2 \kappa B_{\mu \nu}\theta^{\nu \rho} \pi_\rho - 4\kappa^2 B_{\mu \rho} \theta^{\rho \sigma} B_{\sigma \nu}x^{\prime \nu}) \\ \notag
&=& \int d\sigma (\pi_\mu (G_E^{-1}G)^\mu_{\ \nu} \xi^\nu + 4 \kappa^2 x^{\prime \mu} (B \theta B)_{\mu \nu} \xi^\nu ) = \int d\sigma (\pi_\mu \tilde{\xi}^\mu + \kappa x^{\prime \mu} \tilde{\lambda}_\mu) \, ,
\end{eqnarray}
where
\begin{equation}
\tilde{\xi}^\mu =  (G_E^{-1}G)^\mu_{\ \nu} \xi^\nu \, , \quad \tilde{\lambda}_\mu = 4 \kappa  (B \theta B)_{\mu \nu} \xi^\nu \, .
\end{equation}
Using relations (\ref{eq:thetadef}) and (\ref{eq:decomp}), we have
\begin{eqnarray}
4 \kappa (B \theta B)_{\mu \nu} &=& -8 (B G^{-1} B G^{-1}_E B G^{-1}G)_{\mu \nu} = -8 (B G^{-1} B G^{-1} B G_E^{-1}G)_{\mu \nu} \\ \notag
&=& -8 (B^{3} G_E^{-1}G)_{\mu \nu} \, ,
\end{eqnarray}
where $B^{3}$ is defined as in (\ref{eq:B-cube}). From the previous relation we can easily obtain the expression connecting two parameters
\begin{equation}
\tilde{\lambda}_\mu = (B^{3})_{\mu \nu}  \tilde{\xi}^\nu \, .
\end{equation}
We know from the Chapter 4 that this will be the Dirac structure only for $d (B^{3})= 0$, i.e.
\begin{equation}
\Big[ {\cal I}_{B} (\Lambda_1),  {\cal I}_{B} (\Lambda_2) \Big]_{{\cal C}_{B, \theta}} = {\cal I}_{B} \Big( [\Lambda_1, \Lambda_2]_{{\cal C}_{B, \theta}} \Big) \, , \quad d (B^{3})= 0 \, .
\end{equation}

On the sub-bundle ${\cal I}_{\theta}$, the generator (\ref{eq:GBth}) becomes
\begin{equation}
{\cal G}^{B, \theta}_{{\cal I}_{\theta}(\Lambda)} = \int d\sigma \kappa x^{\prime \mu} \lambda_\mu \, ,
\end{equation}
which does not depend on momenta, and therefore we have
\begin{equation}
\Big[ {\cal I}_{\theta} (\Lambda_1),  {\cal I}_{\theta} (\Lambda_2) \Big]_{{\cal C}_{B, \theta}} = 0 \, .
\end{equation}
After the successive twisting by $B$ and then by $\theta$, for sub-bundle ${\cal I}_B$ to be a Dirac structure, the constraints on background fields have to be imposed, while ${\cal I}_{\theta}$ is always the Dirac structure. It is interesting to see what happens when both $B$ and $\theta$ are introduced simultaneously. This is precisely what we analyze in the next chapter.

\section{Courant bracket simultaneously twisted by both B and $\theta$}
\cleq

The matrices $e^{\hat{B}}$ (\ref{eq:ebhat}) and $e^{\hat{\theta}}$ (\ref{eq:enateta}) are not commutative, indicating that the successive twisting inherently treats the fields $B$ and $\theta$ in a different way. In our previous work \cite{CBTh}, the Courant bracket was simultaneously twisted by a 2-form $B$ and a bi-vector $\theta$ in an attempt to derive a twisted Courant bracket that encompasses all forms of generalized fluxes while at the same time being invariant under the self T-duality. The idea was to consider the twist by the matrix $e^{\breve{B}}$, where
\begin{equation} \label{eq:breve}
\breve{B} = \hat{B}+\hat{\theta} = 
\begin{pmatrix}
0 & \kappa \theta^{\mu \nu} \\
2 B_{\mu \nu} & 0
\end{pmatrix} \, ,
\end{equation}
which is constructed to be invariant under T-duality. The expression for $e^{\breve{B}}$ requires some work, as it is not the case that $\breve{B}^2 = 0$ (as in previous examples). Consequentially, it is necessary to obtain all terms in the Taylor expansion of $e^{\breve{B}}$. The expression for $e^{\breve{B}}$  \footnote{The reader is referred to \cite{CBTh} for a step by step derivation of the expression (\ref{eq:ebb}).} is given by
\begin{equation} \label{eq:ebb}
e^{\breve{B}} = 
\begin{pmatrix}
{\cal C}^{\mu}_{\ \nu} & \kappa {\cal S}^\mu_{\ \rho} \theta^{\rho \nu} \\
2 B_{\mu \rho} {\cal S}^{\rho}_{\ \nu} & ( {\cal C}^T)^{\ \nu}_{\mu}
\end{pmatrix} \, ,
\end{equation}
where ${\cal C}$ and ${\cal S}$ are hyperbolic functions of both $B$ and $\theta$, defined by
\begin{equation} \label{eq:calCSdef}
{\cal C}^\mu_{\ \nu} = \cosh{\sqrt{\alpha}}^\mu_{\ \nu} \, , \quad {\cal S}^\mu_{\ \nu} = \Big(\frac{\sinh{\sqrt{\alpha}}}{\sqrt{\alpha}} \Big)^\mu_{\ \nu} \, ,
\end{equation}
where $\alpha^\mu_{\ \nu} = 2\kappa \theta^{\mu \rho} B_{\rho \nu}$. 

The Courant bracket twisted at the same time by $B$ and $\theta$ can be obtained from the Poisson bracket algebra of the generator
\begin{equation} \label{eq:breveG}
\breve{\cal{G}}_{\breve{\Lambda}}= \int d\sigma \Big[ \xi^\mu \breve{\iota}_\mu + \lambda_\mu \breve{k}^\mu \Big] = \int d\sigma \langle  \Lambda , \breve{X}  \rangle \, ,
\end{equation}
where
\begin{equation} \label{eq:breveX}
\breve{X}^M = \begin{pmatrix}
\breve{k}^\mu \\
\breve{\iota}_\mu
\end{pmatrix} = 
\begin{pmatrix}
\kappa {\cal C}^\mu_{\ \nu} x^{\prime \nu} + \kappa ({\cal S} \theta)^{\mu \nu} \pi_\nu \\
2  (B{\cal S})_{\mu \nu} x^{\prime \nu} + ( {\cal C}^T)^{\ \nu}_{\mu} \pi_\nu
\end{pmatrix} \, .
\end{equation}
The algebra of these currents closes with generalized fluxes as the structure functions. Specifically, we have
\begin{equation}
\{\breve{X}^M , \breve{X}^N \} = -\breve{F}^{MN}_{\ \ \ \ P}\ \breve{X}^P \delta(\sigma-\bar{\sigma}) + \kappa \eta^{MN} \delta^{\prime}(\sigma-\bar{\sigma}) \, ,
\end{equation}
where
\begin{eqnarray}\label{eq:breveFdef}
 \breve{F}^{M N \rho} =
\begin{pmatrix}
  \kappa^2  \breve{{\cal R}}^{\mu \nu \rho}  & - \kappa \breve{{\cal Q}}_\nu^{\ \mu \rho} \\
\kappa \breve{{\cal Q}}_\mu^{\ \nu \rho}   &  \breve{{\cal F}}^{\ \rho}_{\!\!\mu \nu}  \\
\end{pmatrix}  \,  , \qquad
 \breve{F}^{M N}{}_\rho  =
\begin{pmatrix}
  \kappa \breve{{\cal Q}}_\rho^{\ \mu \nu}   &  \breve{{\cal F}}^{\ \mu}_{\!\!\nu \rho }  \\
 -\breve{{\cal F}}^{\ \nu}_{\!\!\mu \rho}   &   2 {\cal \breve{B}}_{\mu \nu \rho} \\
\end{pmatrix} \, .
 \end{eqnarray}
The expressions for generalized fluxes were derived in \cite{fluksevi-rad}. Starting with the generalized $\cal{B}_{\mu \nu \rho}$ flux, it is given by
\begin{equation} \label{eq:breveB}
\breve{\cal B}_{\mu \nu \rho} = {\cal C}^\alpha_{\ \mu}  {\cal C}^\beta_{\ \nu}  {\cal C}^\gamma_{\ \rho} (\partial_{\alpha}(B{\cal S}{\cal C}^{-1})_{\beta \gamma} + \partial_{\beta} (B{\cal S}{\cal C}^{-1})_{\gamma \alpha} +\partial_{\gamma}(B{\cal S}{\cal C}^{-1})_{\alpha \beta} ) \, ,
\end{equation}
which can be significantly simplified once one introduces the Lie algebroid with the matrix ${\cal C}$ (\ref{eq:calCSdef}) acting by simple matrix multiplication as its anchor 
\begin{equation}
{\cal C} [\xi_1, \xi_2]_{\hat{L}} = [{\cal C}\xi_1, {\cal C}\xi_2]_L \, , 
\end{equation}
where by $[\xi_1, \xi_2]_{\hat{L}}$ we have marked the twisted Lie bracket, given by
\begin{equation} \label{eq:twisted-Lie-def}
[\xi_1, \xi_2]_{\hat{L}} = {\cal C}^{-1}[{\cal C}\xi_1, {\cal C}\xi_2]_L  \, .
\end{equation}
It has been proven \cite{fluksevi-rad} that the generalized ${\cal B}_{\mu \nu \rho}$ flux can be written simply as
\begin{equation}
\breve{B}_{\mu \nu \rho} = (\hat{d} \hat{B})_{\mu \nu \rho} \, , 
\end{equation}
where $\hat{B}_{\mu \nu} = (B {\cal S} {\cal C})_{\mu \nu}$, and $\hat{d}$ is the exterior derivative associated with the twisted Lie bracket (\ref{eq:twisted-Lie-def}) via (\ref{eq:ext-der-rho}).

As for the generalized ${\cal \breve{F}}$-flux, its expression is given by
\begin{equation} \label{eq:breveFF}
\breve{\cal F}_{\!\!\mu \nu}^{\ \rho} = \breve{f}_{\!\mu \nu}^{\ \rho} - 2\kappa \breve{{\cal B}}_{\mu \nu \sigma} \breve{\theta}^{\sigma \rho} \, , 
\end{equation}
where
\begin{equation} \label{eq:brevef}
\breve{f}_{\!\mu \nu}^{\ \rho}  =  ({\cal C}^{-1})^{\rho}_{\ \sigma} \Big( \hat{\partial}_\mu{\cal C}^{\sigma}_{\ \nu}-   \hat{\partial}_\nu {\cal C}^{\sigma}_{\ \mu}\Big) \, ,
\end{equation} 
and 
\begin{equation}
\hat{\partial}_\mu = ({\cal C}^T)_\mu^{\ \nu} \partial_\nu \, .
\end{equation}
The $\breve{f}$-flux resembles the usually denoted ''geometric'' $f$-flux, which is the consequence of using the non-holonomic basis. However, here we have a more complicated situation, as this appears in a holonomic basis. Moreover, this flux depends on both $B$ and $\theta$. It can be simply obtained from the twisted Lie bracket of partial derivatives
\begin{equation} \label{eq:f-twisted-def}
[\partial_\mu, \partial_\nu]_{\hat{L}}= \breve{f}_{\!\mu \nu}^{\ \rho}\ \partial_\rho \, .
\end{equation}

As for the generalized ${\cal \breve{Q}}$-flux, we have
\begin{equation} \label{eq:breveQQ}
\breve{\cal Q}_\rho^{\ \mu \nu} = \breve{Q}_\rho^{\ \mu \nu} + 2\kappa \breve{\theta}^{\mu \alpha} \breve{\theta}^{\nu \beta} {\cal \breve{B}}_{\rho \alpha \beta } \, ,
\end{equation}
where
\begin{eqnarray} \label{eq:Qbreve}
\breve{Q}_\rho^{\ \mu \nu} =\hat{\partial}_\rho \breve{\theta}^{\mu \nu} + \breve{f}^{\ \mu}_{\!\rho \sigma}\  \breve{\theta}^{\sigma \nu} - \breve{f}^{\ \nu}_{\!\rho \sigma}\ \breve{\theta}^{\sigma \mu} \, ,
\end{eqnarray}
where
\begin{equation} \label{eq:brthdef}
\breve{\theta}^{\mu \nu} = ({\cal S C}^{-1})^\mu_{\ \rho} \theta^{ \rho\nu} \, .
\end{equation}
Substituting the twisted counterparts of the Lie bracket into expression for the Koszul bracket, one can introduce the twisted Koszul bracket, given by
\begin{equation} \label{eq:twist-koszul}
[\lambda_1, \lambda_2]_{\breve{\theta}} = {\cal \hat{L}}_{\breve{\theta}(\lambda_1)} \lambda_2- {\cal \hat{L}}_{\breve{\theta}(\lambda_2)} \lambda_1 - \hat{d}(\breve{\theta}(\lambda_1, \lambda_2)) \, .
\end{equation}
Taking the twisted Koszul bracket of two trivial one-forms, one obtains the $\breve{Q}$-flux
\begin{equation} \label{eq:Q-twisted-def}
[dx^\mu, dx^\nu]_{\breve{\theta}} = \breve{Q}_\rho^{\ \mu \nu}\ dx^\rho \, .
\end{equation}

Lastly, the generalized ${\cal \breve{R}}$-flux is given by
\begin{equation} \label{eq:breveRR}
\breve{\cal R}^{\mu \nu \rho} = \breve{R}^{\mu \nu \rho} + 2\kappa \breve{\theta}^{\mu \alpha} \breve{\theta}^{\nu \beta} \breve{\theta}^{\rho \gamma} {\cal \breve{B}}_{\alpha \beta \gamma} \, ,
\end{equation}
where
 \begin{eqnarray} \label{eq:Rbreve}
\breve{R}^{\mu \nu \rho} = \breve{\theta}^{\mu \sigma} \hat{\partial}_\sigma \breve{\theta}^{\nu \rho} + \breve{\theta}^{\nu \sigma} \hat{\partial}_\sigma \breve{\theta}^{\rho \mu}+ \breve{\theta}^{\rho \sigma} \hat{\partial}_\sigma \breve{\theta}^{\mu \nu}- (\breve{\theta}^{\mu \alpha} \breve{\theta}^{\rho \beta} \breve{f}_{\!\alpha \beta}^{\ \nu} + \breve{\theta}^{\nu \alpha} \breve{\theta}^{\mu \beta} \breve{f}_{\!\alpha \beta}^{\ \rho}+\breve{\theta}^{\rho \alpha} \breve{\theta}^{\nu \beta} \breve{f}_{\!\alpha \beta}^{\ \mu}) \, .
\end{eqnarray}
We can make some sense of it if we consider the twisted Schouten-Nijenhuis bracket, as the generalization of the twisted Lie bracket on the space of multi-vectors
\begin{eqnarray}
[f, g]_{\hat{S}} = 0 \, , \quad [\xi, f]_{\hat{S}} = {\cal \hat{L}}_{\xi}(f) \, , \quad [\xi_1, \xi_2]_{\hat{S}} = [\xi_1, \xi_2]_{\hat{L}} \, ,
\end{eqnarray}
with other brackets defined using the appropriate graded identities. Then, it has been shown \cite{fluksevi-rad} that
\begin{equation} \label{eq:breveR-thth}
\breve{R}=\frac{1}{2}[\breve{\theta}, \breve{\theta}]_{\hat{S}}  \, .
\end{equation}
We see that in case of the simultaneous twisting, all generalized fluxes exist. They necessitate some other twisted structures to be defined and all of them depend on both fields $B$ and $\theta$.

 
\subsection{Courant algebroid}

The Courant bracket twisted at the same time by $B$ and $\theta$ defines the Courant algebroid with 
\begin{equation}
\rho_{\breve{B}} \Lambda = \pi   e^{-\breve{B}} \Lambda = \begin{pmatrix}
{\cal C}\xi  - \kappa ({\cal S}\theta) \lambda \\
0
\end{pmatrix}\, ,
\end{equation}
as its anchor, and
\begin{equation}
{\cal D}_{\breve{B}}f = e^{\breve{B}}  d f = \begin{pmatrix}
\kappa {\cal S}\theta\ d f  \\
{\cal C}^T d f
\end{pmatrix} \, 
\end{equation}
as its differential operator.

\subsection{Dirac structures}

Consider following fields 
\begin{equation} \label{eq:brBTh}
 \breve{B} =  B {\cal S}{\cal C}^{-1} \, , \quad \breve{\theta} =   {\cal S}{\cal C}^{-1}\theta \, .
\end{equation}
We can define the graph over tangent bundle as 
\begin{equation} \label{eq:breveib}
{\cal I}_{\breve{B}} (\Lambda) = \xi^\mu \oplus 2\breve{B}_{\mu \nu} \xi^\nu \, ,
\end{equation}
and over cotangent bundle as
\begin{equation}\label{eq:breveith}
{\cal I}_{\breve{\theta}} (\Lambda) = \breve{\theta}^{\mu \nu} \lambda_\nu \oplus \lambda_\mu \, .
\end{equation}

In the former case, the generator $\breve{G}$ becomes
\begin{eqnarray} \label{eq:bg1}
\breve{\cal{G}}_{{\cal I}_{\breve{B}}(\breve{\Lambda}) }&=& \int d\sigma \Big(\breve{\xi}^\mu(\pi_\nu {\cal C}^\nu_{\ \mu} + 2 \breve{B}_{\mu \rho} {\cal C}^{\rho}_{\ \nu} \kappa x^{\prime \nu})+2 \breve{B}_{\mu \nu} \breve{\xi}^\nu ({\cal C}^\mu_{\ \rho} \kappa x^{\prime \rho} + \kappa {\cal C}^\mu_{\ \rho} \breve{\theta}^{\rho \nu} \pi_\nu) \Big) \\ \notag
&=& \int d\sigma \pi_\mu ({\cal C}^{-1})^\mu_{\ \nu} \breve{\xi}^\nu \, ,
\end{eqnarray}
where we have used the well known identity between hyperbolic functions
\begin{equation}
1 = \cosh^2 \sqrt{\alpha} - \sinh^2 \sqrt{\alpha}= {\cal C}^2 - 2\kappa \breve{\theta} \breve{B} {\cal C}^2 \, .
\end{equation}
The generator (\ref{eq:bg1}) is just the generator of the general coordinate transformations. Since such a generator gives rise to the Lie bracket, the sub-bundle ${\cal I}_{\breve{B}}$ (\ref{eq:breveib}) is Dirac structure.

In the latter case, the generator  $\breve{{\cal G}}$ becomes
\begin{eqnarray} \label{eq:bg2}
\breve{\cal{G}}_{{\cal I}_{\breve{\theta}}(\breve{\Lambda})}&=& \int d\sigma \Big(\breve{\theta}^{\mu\nu} \breve{\lambda}_\nu (\pi_\rho {\cal C}^\rho_{\ \mu} + 2 \breve{B}_{\mu \rho} {\cal C}^{\rho}_{\ \sigma} \kappa x^{\prime \sigma})+2 \breve{\lambda}_{\mu} ({\cal C}^\mu_{\ \rho} \kappa x^{\prime \rho} + \kappa {\cal C}^\mu_{\ \rho} \breve{\theta}^{\rho \nu} \pi_\nu) \Big) \\ \notag
&=& \int d\sigma  ({\cal C}^{-1})^\mu_{\ \nu} x^{\prime \nu} \breve{\lambda}_\mu \, .
\end{eqnarray}
However, since there is no dependence on momenta, the Poisson bracket algebra of this generator is zero. Hence, we obtain 
\begin{equation}
[{\cal I}_{\breve{\theta}}(\Lambda_1),  {\cal I}_{\breve{\theta}}(\Lambda_2)]_{{\cal C}_{\theta B}}  = 0 \, .
\end{equation}
Again, we see this is a Dirac structure for all $\breve{\theta}$. This is a very interesting result - we were able to construct the Courant algebroid invariant under T-duality, that contains all generalized fluxes, such that they can exist on its Dirac structures without any constraints put on them.

\section{Conclusion}
\cleq

The idea of this paper was to research and describe the family of Courant algebroids that are generated by the elements of the $O(D,D)$ group, with the focus on Dirac structures for these algebroids. Every $O(D,D)$ transformation acts as an isomorphism between the standard Courant bracket and the appropriate twisted Courant bracket. We showed how the twisted bracket, together with the $O(D,D)$ invariant inner product and the anchor obtained from the composition of the inverse of the corresponding $O(D,D)$ transformation and the projection on the tangent bundle defines the Courant algebroid, where all compatibility conditions are a priori satisfied. 

The first twisted Courant algebroid considered was the one defined with the Courant bracket twisted by a 2-form $B$. This is the most well known example of a deformation of the Courant bracket, that results in additional term related to the $H$-flux. We demonstrated that the sub-bundle ${\cal I}_B$ (\ref{eq:isoB}) is the Dirac structure regardless of whether the 2-form $B$ is closed. On the other hand, the sub-bundle ${\cal I}_{\theta}$ (\ref{eq:isoth}) is a Dirac structure only if the generalized $R$ flux is zero. 

The second example considered was the Courant algebroid with the $\theta$-twisted Courant bracket as its defining bracket. Again, we investigated the Dirac structures in the form  ${\cal I}_B$  and ${\cal I}_{\theta}$. In the former case we obtained that additional constraint on fluxes needs to be imposed in order for it to be Dirac structure. We showed that the latter case is the Dirac structure regardless whether the bi-vector $\theta$ is Poisson. Moreover, we showed that the bracket on such Dirac structure reduces to the Courant bracket of two pure 1-forms and is therefore zero. 

The subsequent example was the case of Courant algebroid defined with the Courant bracket twisted by $B$ and then by $\theta$, also known as the Roytenberg bracket. Even though we saw that the twisting by $B$ allowed for existence of arbitrary $H$-flux on Dirac structures, when that bracket is further twisted by $\theta$, another constraints on $H$ flux on its Dirac structures appear.

The last example is the one of twisting the Courant bracket simultaneously by $B$ and $\theta$. We showed that such a bracket features all generalized fluxes ${\cal \breve{H}},\  {\cal \breve{F}},\ {\cal \breve{Q}},\ {\cal \breve{R}}$, that are themselves functions of some effective fields $\breve{B}$ and $\breve{\theta}$, which are both functions of both $B$ and $\theta$. We were then looking for Dirac structures in the form ${\cal I}_{\breve{B}}$ and ${\cal I}_{\breve{\theta}}$ and showed that no conditions on the effective background fields is necessary for such sub-bundles to be Dirac structures.
\begin{table}
\begin{center}
\begin{tabular}{|c|c|c|c|} 
\hline
{\bf Bracket }& {\bf Transformation} &{\bf Anchor }& {\bf Derivative}\\
\hline
\tiny{$[\Lambda_1, \Lambda_2]_{\cal C}$} & - & \tiny{$\pi = \begin{pmatrix}
1 & 0 \\
0 & 0 
\end{pmatrix}$} & \tiny{$d f$}
\\
\hline 
\tiny{$e^T[e^{-T}\Lambda_1, e^{-T}\Lambda_2]_{\cal C}$} & \tiny{$e^{T},\quad T = 
\begin{pmatrix}
A & \theta \\
B & -A^T 
\end{pmatrix} $} & \tiny{$\rho_T = e^{-T} \pi$} & \tiny{$e^T d f$}
\\
\hline 
\tiny{$e^{\hat{B}} [e^{-\hat{B}}\Lambda_1, e^{-\hat{B}}\Lambda_2]_{\cal C}$} &\tiny{$e^{\hat{B}} = \begin{pmatrix}
1 & 0 \\
2B & 1
\end{pmatrix}$} & \tiny{$\pi = \begin{pmatrix}
1 & 0 \\
0 & 0 
\end{pmatrix} $} & \tiny{$d f$}
\\
\hline 
\tiny{$e^{\hat{\theta}} [e^{-\hat{\theta}}\Lambda_1, e^{-\hat{\theta}}\Lambda_2]_{\cal C}$} & \tiny{$e^{\hat{\theta}} = 
\begin{pmatrix}
1 & \kappa \theta \\
0 & 1
\end{pmatrix}$} & \tiny{$\rho_{\theta}  =
\begin{pmatrix}
1 & -\kappa \theta \\
0 & 0 
\end{pmatrix}$} & \tiny{${\cal D}_{\theta}f = \begin{pmatrix}
\kappa \theta d f \\
d f
\end{pmatrix}$}
\\
\hline 
\tiny{$ e^{\hat{\theta}} e^{\hat{B}} [e^{-\hat{B}}e^{-\hat{\theta}}\Lambda_1, e^{-\hat{B}}e^{-\hat{\theta}} \Lambda_2]_{\cal C}$} & \tiny{$e^{\hat{\theta}} e^{\hat{B}} = \begin{pmatrix}
1 + 2\kappa \theta B & \kappa \theta \\
2 B & 1 
\end{pmatrix}$} & \tiny{$\rho_{B, \theta} = 
\begin{pmatrix}
1 & -\kappa \theta \\
0 & 0 
\end{pmatrix}$} & \tiny{${\cal D}_{B, \theta}f = \begin{pmatrix}
\kappa \theta d f \\
d f
\end{pmatrix}$}
\\
\hline 
& \tiny{$e^{\breve{B}} = 
\begin{pmatrix}
{\cal C}^{\mu}_{\ \nu} & \kappa {\cal S}^\mu_{\ \rho} \theta^{\rho \nu} \\
2 B_{\mu \rho} {\cal S}^{\rho}_{\ \nu} & ( {\cal C}^T)^{\ \nu}_{\mu}
\end{pmatrix}$} & &
\\
\tiny{$e^{\breve{B}}[e^{-\breve{B}}\Lambda_1, e^{-\breve{B}} \Lambda_2]_{\cal C}$}  &  \tiny{${\cal C}^\mu_{\ \nu} = \cosh{\sqrt{\alpha}}^\mu_{\ \nu} \, , \ {\cal S}^\mu_{\ \nu} = \Big(\frac{\sinh{\sqrt{\alpha}}}{\sqrt{\alpha}} \Big)^\mu_{\ \nu} $} & \tiny{$\rho_{\breve{B}} = \begin{pmatrix}
{\cal C} & -\kappa {\cal S}\theta \\
0 & 0 
\end{pmatrix} $} & \tiny{${\cal D}_{\breve{B}}f  = \begin{pmatrix}
\kappa {\cal S}\theta\ d f  \\
{\cal C}^T d f
\end{pmatrix}$}
\\
&  \tiny{$\alpha^\mu_{\ \nu} = 2\kappa \theta^{\mu \rho}B_{\rho \nu} $} & &
\\
\hline 
\end{tabular}
\caption{Overview of Courant algebroids obtained in the paper}
\end{center}
\end{table}

The Dirac structures for different Courant algebroids were already investigated in \cite{nilm} in search of the possible coexistence of fluxes on nilmanifolds. There, the authors considered the Courant bracket, as well as different twisted Courant brackets and Dirac structures on their respective algebroids. There are a few differences in our approach, that we would like explicitly to outline.

Firstly, we use a different definition for the twisted Courant bracket. There are few terms that are widely circulating and having ambiguous meaning in physics as the term ''twisted''. The authors in \cite{nilm} define a $H$-twisted and $R$-twisted Courant bracket as the ordinary bracket with one additional term containing the $H$-flux and $R$-flux respectively. We define the twisted Courant brackets by the underlining $O(D, D)$ transformation that governs the twist by the expression (\ref{eq:twdef}). Using the former definition, the $H$-twisted Courant bracket correspond to the $B$-twisted Courant bracket using the latter definition. However, $R$-twisted Courant bracket does not correspond to any bracket defined via expression (\ref{eq:twdef}) - the $\theta$-twisted Courant bracket contains other terms, not only the one related to the $R$-flux.

Secondly, the brackets that we obtained are by construction the brackets that define the Courant algebroids, as proven in the Chapter 3 of the paper. This is not the case for all brackets in \cite{nilm}, and the authors resolve it by applying the additional $O(D, D)$ rotation to the deformed brackets.

Thirdly, we obtain different constraints imposed on fluxes on Dirac structures. Though they are the same conditions in case of twisting by a 2-form $B$, they are less restrictive in our approach in case of twisting by $\theta$ and in case of twisting by $B$ and $\theta$. In addition, we obtain the Courant algebroid that is twisted with all fluxes, while these fluxes can coexist on Dirac structures.

\begin{table}
\begin{center}
\begin{tabular}{|c|c|c|}
\hline
{\bf Bracket} & {\bf $ {\cal I}_B (\Lambda) = \xi^\mu \oplus 2 B_{\mu \nu} \xi^\nu $}& $ {\cal I}_{\theta} (\Lambda) = \kappa \theta^{\mu \nu} \lambda_\nu \oplus \lambda_\mu$\\
\hline
\tiny{$[\Lambda_1, \Lambda_2]_{\cal C}$} & \tiny{$H = d B = 0$} & \tiny{ $ R=[\theta, \theta]_S = 0$} 
\\
\hline
\tiny{$e^{\hat{B}} [e^{-\hat{B}}\Lambda_1, e^{-\hat{B}}\Lambda_2]_{\cal C}, \quad \hat{B} = \begin{pmatrix}
0 & 0 \\
2B & 0
\end{pmatrix}$} &\tiny{$\forall \ H = d B$} & \tiny{${\cal R}=[\theta, \theta]_S - \wedge^3 \theta \ d B =0$}
\\
\hline
\tiny{$e^{\hat{\theta}} [e^{-\hat{\theta}}\Lambda_1, e^{-\hat{\theta}}\Lambda_2]_{\cal C}, \quad \hat{\theta} = \begin{pmatrix}
0 & \kappa \theta \\
0 & 0
\end{pmatrix}$}   & \tiny{$dB -4 d (B^3) = 0$} & \tiny{$\forall \ R$} \\
\hline
\tiny{$ e^{\hat{\theta}} e^{\hat{B}} [e^{-\hat{B}}e^{-\hat{\theta}}\Lambda_1, e^{-\hat{B}}e^{-\hat{\theta}} \Lambda_2]_{\cal C}$} & \tiny{$d(B^{3}) = 0$}  & \tiny{$\forall \ R$} \\
\hline
\tiny{$e^{\breve{B}}[e^{-\breve{B}}\Lambda_1, e^{-\breve{B}} \Lambda_2]_{\cal C}, \quad \breve{B} = \hat{B} + \hat{\theta} = \begin{pmatrix}
0 & \kappa \theta \\
2B & 0
\end{pmatrix}$} & \tiny{$ \forall \ \breve{B} = B \coth{\sqrt{2 \kappa \theta B}}$} & \tiny{$\forall \ \breve{\theta} = \coth{\sqrt{2 \kappa \theta B\theta}}$}\\
 & \tiny{${\cal I}_{\breve{B}} (\Lambda) = \xi^\mu \oplus 2\breve{B}_{\mu \nu} \xi^\nu $} & \tiny{${\cal I}_{\breve{\theta}} (\Lambda) = \kappa \breve{\theta}^{\mu \nu} \lambda_\nu \oplus \lambda_\mu$}\\
\hline
\end{tabular}
\end{center}
\caption{Dirac structures corresponding to different Courant algebroid's brackets}
\end{table}

The importance of the Dirac structures for physicist among others reflects in their interpretation as the D-branes in string theory. The D-branes represent the class of extended dynamical objects upon which open strings end with Dirichlet boundary conditions being imposed on its ends. They can be interpreted as leaves in a foliation, where the foliation deformations are interpreted as the scalar and vector field fluctuations of the D-branes. The transformations governed by these fluctuations is closed in the space of the Dirac structures \cite{asakawa}. As such, the results of this paper have a potential to further understand the fluxes on D-branes.

\vspace{2 cm}
\textbf{Acknowledgements}
\vspace{.5 cm}

We thank Ljubica Davidović for discussion. This work is supported in part by the Serbian Ministry of Science, and the Institute of Physics Belgrade.


\begin{thebibliography}{}
\bibitem{gualtieri} M. Gualtieri, \textit{Generalized complex geometry} (2003), arXiv:math/0401221.
\bibitem{GCY} N. Hitchin, \textit{Generalized Calabi-Yau manifolds, Quart.J.Math.Oxford Ser.} \textbf{54} (2003) 281-308.
\bibitem{tdual} E. Alvarez, L. Alvarez-Gaume and Y. Lozano, \textit{An introduction to T-duality in string theory, Nucl. Phys. Proc. Suppl.} \textbf{41} (1995) 1-20.
\bibitem{tdual1} A. Giveon, M. Parrati and E. Rabinovici, \textit{Target space duality in string theory, Phys. Rep.} \textbf{244} (1994) 77-202.
\bibitem{tdual2} Y. Lozano, \textit{Duality and canonical transformations, Mod. Phys. Lett.} \textbf{A11} (1996) 2893-2914.
\bibitem{buscher} T. Buscher, \textit{A symmetry of the string background field equations, Phys. Lett.} \textbf{ B 194 }(1987) 51.
\bibitem{buscher1} T. Buscher, \textit{Path-integral derivation of quantum duality in nonlinear sigma-models, Phys. Lett.} \textbf{201} (1988) 466.
\bibitem{doucou} C. Hull, B. Zwiebach, \textit{The gauge algebra of double field theory and Courant brackets, JHEP} \textbf{01} (2015) 012.
\bibitem{dualsim} Lj. Davidovi\' c, B. Sazdovi\' c, \textit{The T-dual symmetries of a bosonic string, Eur. Phys. J.} \textbf{C 78} (2018) 600.
\bibitem{royt} D. Roytenberg, \textit{Quasi-Lie bialgebroids and twisted Poisson manifolds, Letters in Mathematical Physics} \textbf{61} (2002) 123.
\bibitem{royt1} D. Roytenberg, \textit{Courant algebroids, derived brackets and even symplectic supermanifolds} arXiv:math/9910078.
\bibitem{cdual} Lj. Davidovi\' c, I. Ivani\v sevi\' c, B. Sazdovi\' c, \textit{Courant bracket as T-dual invariant extension of Lie bracket, JHEP} \textbf{03} (2021) 109.
\bibitem{CBTh} Lj. Davidovi\' c, I. Ivani\v sevi\' c, B. Sazdovi\' c, \textit{Courant bracket twisted both by a 2-form $B$ and by a bi-vector $\theta$, Eur. Phys. J} \textbf{C 81} (2021) 685.
\bibitem{flux1} E. Plauschinn, \textit{Non-geometric backgrounds in string theory, Phys.Rept.} \textbf{798} (2019) 1-122.
\bibitem{flux2} M. Grana, R. Minasian, M. Petrini, D. Waldram \textit{T-duality, Generalized Geometry and Non-Geometric Backgrounds, JHEP} \textbf{04} (2009) 075.
\bibitem{flux3} R. Blumenhagen, A. Deser, E. Plauschinn, F. Rennecke, \textit{Bianchi Identities for Non-Geometric Fluxes - From Quasi-Poisson Structures to Courant Algebroids, Fortsch. Phys. } \textbf{60} (2012) 1217-1228.
\bibitem{NC-string} D. Lüst, \textit{T-duality and closed string non-commutative (doubled) geometry', JHEP} \textbf{12} (2010) 084.
\bibitem{NA-string}  R. Blumenhagen, A. Deser, D. Lüst, E. Plauschinn and F. Rennecke, \textit{Non-geometric Fluxes, Asymmetric Strings and Nonassociative Geometry, J. Phys.} \textbf{A44} (2011) 385401.
\bibitem{fluxNO} B. Nikolić, D. Obrić, \textit{Noncommutativity and nonassociativity of closed bosonic string on T-dual toroidal backgrounds, Fortschritte der Physik} (2018) 1800009.
\bibitem{aksz} D. Roytenberg, \textit{AKSZ-BV Formalism and Courant Algebroid-induced Topological Field Theories, Lett. Math. Phys.} \textbf{79} (2007) 143-159.
\bibitem{L-infty} C. J. Grewcoe, L. Jonke, \textit{Courant sigma model and $L_{\infty}$-algebras, Fortsch. Phys. } \textbf{68} (2020) 6.
\bibitem{courant} T. Courant, \textit{Dirac manifolds, Trans. Amer. Math. Soc.} \textbf{319} (1990) 631-661.
\bibitem{dirak} T. Courant and A. Weinstein, \textit{Beyond Poisson structures. In Action hamiltoniennes de groupes. Troisi\`eme th\'eor\`eme de Lie (Lyon, 1986), Vol. 27 of Travaux en Cours, Hermann, Paris, 1988} (1988) 39–49.
\bibitem{dirak1} H. Bursztyn, \textit{A brief introduction to Dirac manifolds, lectures at the school on Geometric and Topological Methods for Quantum Field Theory,} arXiv:1112.5037  (2009).
\bibitem{dirak2} I. Dorfman, \textit{Dirac structures of integrable evolution equations, Phys. Lett.} \textbf{A 125} (1993) 240-246.
\bibitem{asakawa} T. Asakawa, S. Sasa, S. Watamura, \textit{D-branes in Generalized Geometry and Dirac-Born-Infeld Action, JHEP} \textbf{10} (2012) 064.
\bibitem{nilm} A. Chatzistavrakidis, L. Jonke, O. Lechtenfeld, \textit{Dirac structures on nilmanifolds and coexistence of fluxes, Nucl. Phys.} \textbf{B 883} (2014) 59-82.
\bibitem{bulk} A. Chatzistavrakidis, L. Jonke, O. Lechtenfeld, \textit{Sigma models for genuinely non-geometric backgrounds, JHEP} \textbf{11} (2015) 182.
\bibitem{courant1} Z.-J. Liu, A. Weinstein and P. Xu, \textit{Manin triples for Lie bialgebroids, J. Differential Geom.} \textbf{45} (1997), 547–574.
\bibitem{crdual} I. Ivani\v sevi\' c, Lj. Davidovi\' c, B. Sazdovi\' c, {\it Courant bracket found out to be T-dual to Roytenberg one, Eur. Phys. J.} {\textbf C 80}, (2020) 571.
\bibitem{fluksevi-rad} Lj. Davidović, I. Ivanišević, B. Sazdović, \textit{Fluxes of Courant bracket twisted at the same time by $B$ and $\theta$}, arxiv: 2312.11268.
\bibitem{action} K. Becker, M. Becker and J. Schwarz \textit{String Theory and M-Theory: A Modern Introduction} (Cambridge University Press, Cambridge, 2007).
\bibitem{regal} B. Zwiebach, \textit{A First Course in String Theory}, (Cambridge University Press, Cambridge, 2004).
\bibitem{liealg} J. Pradines \textit{Theorie de Lie pour les groupoides differentiables, Calcul differentiel dans la categorie des groupoides infinitesimaux}  C. R. Acad. Sci. Paris 264: 245–248 (1967).
\bibitem{drinfeld1} V.G. Drinfel'd, \textit{Quantum groups, Proc. ICM,} Berkeley (1986), 789-820.
\bibitem{drinfeld2} V.G. Drinfel'd, \textit{Quasi-Hopf algebras, Leningrad Math. J.} \textbf{2} (1991) 829-860.
\bibitem{liealg1} K. Mackenzie, \textit{General theory of Lie groupoids and Lie algebroids, Cambridge University Press,  London Mathematical Society Lecture Note} \textbf{213} (2005).
\bibitem{drinfeld3} A. Çatal-Özer, K. Doğan, C. Yetişmişoğlu, \textit{Drinfel'd double of bialgebroids for string and M theories: dual calculus framework, JHEP} \textbf{07} (2024) 30. 
\bibitem{LA-geometry} R. Blumenhagen, A. Deser, E. Plauschinn, F. Rennecke, and C. Schmid, \textit{The Intriguing Structure of Non-geometric Frames in String Theory, Fortsch. Phys.} \textbf{61} 893-925 (2013). 
\bibitem{SNB} J. A. de Azcarraga, A. M. Perelomov, J. C. Perez Bueno, \textit{The Schouten-Nijenhuis bracket, cohomology and generalized Poisson structures, J. Phys.} \textbf{A29} (1996) 7993-8110.
\bibitem{koszul} Y. Kosmann-Schwarzbach, \textit{From Poisson algebras to Gerstenhaber algebras, Annales de l'institut Fourier} \textbf{46} (1996) 1243-1274.
\bibitem{letters} P. Ševera, \textit{Letters to Alan Weinstein about Courant algebroids} (2017), arXiv:1707.00265.
\bibitem{nick1} N. Halmagyi, \textit{Non-geometric string backgrounds and worldsheet algebras, JHEP} \textbf{07} (2008) 137.
\bibitem{WZW} C. Klimcik, T. Strobl, \textit{WZW-Poisson manifolds, J.Geom.Phys.} \textbf{43} (2002) 341-344.

\end{thebibliography}
\end{document}